\documentclass[final,authoryear, 5p, times]{elsarticle}

\usepackage{graphics}
\usepackage{subcaption}
\usepackage{graphicx}
\usepackage{tabularx}

\usepackage{amsmath,amsfonts,amssymb,amsthm}
\usepackage{soul}

\usepackage{booktabs}                   
\usepackage{float}

\usepackage{newtxtext}
\usepackage[varvw]{newtxmath}

\begin{document}

\begin{frontmatter}



\title{Modelling and calibration of pair-rule protein patterns in Drosophila embryo: From Even-skipped and Fushi-tarazu
 to Wingless expression networks}

\author{Catarina Dias}
\ead{catarina.reis.dias@tecnico.ulisboa.pt} 

\author{Rui Dil\~ao}
\ead{ruidilao@tecnico.ulisboa.pt}
\address{University of Lisbon, IST, Dep. of Physics, Nonlinear Dynamics Group, 
Av. Rovisco Pais, 1049-001 Lisbon, Portugal}

\date{\today}

\begin{abstract}
We modelled and calibrated the distributions of the seven-stripe patterns of Even-skipped (\textit{Eve})  and Fushi-tarazu (\textit{Ftz}) pair-rule proteins along the anteroposterior axis of the \textit{Drosphila} embryo, established during early development. We have identified the putative repressive combinations for five \textit{Eve} enhancers, and we have explored the relationship between \textit{Eve} and  \textit{Ftz} for complementary patterns. The regulators of \textit{Eve} and \textit{Ftz} are stripe-specific DNA enhancers with embryo position-dependent activation rates and are regulated by the gap family of proteins.  We achieved remarkable data matching of the \textit{Eve} stripe pattern, and the calibrated model reproduces gap gene mutation experiments.  Extended work inferring the Wingless (\textit{Wg}) fourteen stripe pattern from \textit{Eve} and \textit{Ftz} enhancers have been proposed,  clarifying the hierarchical structure of \textit{Drosphila}'s genetic expression network during early development.
\end{abstract}


\end{frontmatter}

\section{Introduction}
\label{sec:intro}

The emergence of periodic band structures along the anteroposterior axis of the \textit{Drosophila} embryo during early development characterises \textit{Drosophila} embryogenesis, one of the most thoroughly studied gene regulatory networks (\cite{Nuss}). Broad domains of maternal proteins regulate the expression of a set of zygotic genes -- the gap genes, along the anteroposterior axis of the embryo (\cite{akam}, \cite{Riv}, \cite{aldi2006}). At a later stage of development, a new set of proteins form seven-stripe complementary patterns -- pair-rule proteins,  of which the most studied pair is the  Even-skipped  (\textit{Eve}) and  Fushi-tarazu  (\textit{Ftz}) (\cite{Frash_correct}).
This cascade of gene-expressed proteins (maternal $\to$ gap $\to$ pair-rule) terminates with the segment-polarity gene-expressed proteins that double the pattern to fourteen stripes, after which cellularisation begins. Cells in the region of each strip, at a later stage of development, will determine the segmented larvae body plan.

It is presumed that each \textit{Eve} stripe, and possibly each \textit{Ftz} stripe, is formed and controlled individually by a specific set of transcription factors that regulate individual enhancers  (\cite{Nuss}) -- regions of the DNA that affect gene transcription and where transcription factors bind. Over the past decades, \textit{eve's} transcriptional control was studied via mutation experiments.   The most recent depiction of the \textit{eve} locus includes five separate enhancers: two dual stripe elements 3+7 and 4+6, and three single stripe elements 1, 2 and 5 (\cite{Lim2018}). The classical view of the stripe 2 enhancer includes two transcription repressors, Krüppel (\textit{Kr}) and Giant (\textit{Gt}), and two transcription activators, Bicoid (\textit{Bcd}) and Hunchback (\textit{Hb}), who work synergistically (\cite{Small1992},  \cite{SIMPSONBROSE1994855}). However, \cite{Vincent2018HunchbackIC} proposed an alternative hypothesis that \textit{Hb} represses the stripe 2 enhancer and that its repression can be counteracted by nearby sequences, possibly by Caudal (\textit{Cad}), a protein whose mRNA has maternal origin. As for stripe 5, little information is provided, but it is proposed to be regulated by two transcription repressors, \textit{Kr} and \textit{Gt} (\cite{Fuji}). Finally, the two dual stripe enhancers are believed to be regulated by the same transcription repressors, \textit{Hb} and Knirps (\textit{Kni}) (\cite{SMALL1996314}, \cite{Fuji}). In the latter case, it is suggested that the enhancers are sensitive to repression by the number and affinity to repressor-binding sites, implying that the stripes 3+7 and 4+6 enhancers would respond autonomously to different amounts of \textit{Hb} and \textit{Kni} (\cite{Clyde}, \cite{Crocker2016QuantitativelyPC}). It is also suggested that although stripe 7 is not separable from stripe 3, its activation requires sequences outside of the minimal stripe 3 element. This led to the suggestion of an extra enhancer, a 2+7 dual element, which would be an extended version of the minimal \textit{eve} stripe 2 enhancer (\cite{Vincent2018HunchbackIC}). 
 To date, practically nothing is known about the formation of stripe 1, with only a mention that it may be activated by \textit{Hb} (\cite{Fuji}).

Despite the many experiments for \textit{Eve}, there is still no consensus on the minimum set of its regulators or any successful mathematical modelling, calibration, or spatial distribution of the full seven-stripe pattern of \textit{Eve}. It is even less clear how the formation of the \textit{Ftz} pattern occurs. The latest hypothesis (\cite{Schroeder2}) suggests three enhancers initiating stripe pairs 1+5, 2+7, and 3+6, and the   \textit{Zebra} enhancer driving the expression of stripe 4, but, as far as we know, no transcription factors are known or reported in the literature. 

In this paper, we develop a simple kinetic framework that accurately predicts and calibrates the formation of the \textit{Eve} stripe pattern,  providing a minimum set of gap regulators for each \textit{eve} enhancer. We use the mass-action law to describe every biological process describing the expression of \textit{Eve} along the anteroposterior axis of the \textit{Drosophila} embryo. We anchor this work on the data provided by the FlyEx database (\cite{Kozlov}, \cite{Myasnikova}, \cite{Poustelnikova}, \cite{Pisarev}). Based on the success of this modelling approach for \textit{Eve}, we can predict the position of six of the seven stripes of \textit{Ftz}, and we have extended our approach to predict the fourteen stripes of the segment-polarity protein Wingless (Wg).  The nomenclature for gene and protein names follows the convection adapted from \cite[p. ix]{Alber}: the first letter of protein names is uppercase, and the acronymous follows the same rule but is typed in italics. For genes, the first letter is lowercase and typed in italics.

\section{Methods}
\label{sec:imple}

To describe all the genetic regulatory interactions that occur during the embryonic segmentation of \textit{Drosophila}, we adopted the \cite{ALVES2005429} approach. We considered that the kinetic mechanisms describing the non-regulated production of a protein P is
\begin{equation} \label{free-prod}
        \begin{array}{l}\displaystyle 
             \text{G}\mathop{\longrightarrow}^{k_{1}} \text{G} + \text{P} \\ \displaystyle
             \text{P} \mathop{\longrightarrow}^{d}
        \end{array}
\end{equation}
where G is the gene for P. We assume that DNA splicing has occurred,  the transcription and translation from DNA to protein and the concentration of polymerases are represented by the rate constant $k_1$, and that $d$ is the intrinsic degradation rate of protein P.  
The generic kinetic mechanisms that describe gene repression is
    \begin{equation} \label{repression}
      \begin{array}{l}\displaystyle 
            \text{G} \mathop{\longrightarrow}^{k_{1}} \text{G} + \text{P} \\ \displaystyle 
          	\text{R} + \text{G} \mathop{\rightleftarrows}^{k_{2}}_{k_{-2}} \text{G}_{\text{R}} \\ \displaystyle
            \text{P} \mathop{\longrightarrow}^{d} 
         \end{array}
    \end{equation}
where R is a repressor transcription factor, G$_\text{R}$ is a gene complex and $k_i$ are reaction rates.  
In mechanism \eqref{repression}, the regulation of the transcription factor is not considered since we are only interested in the repressed translation states of gene G.

\begin{figure}[h!]
    \centering
    \includegraphics[width=\columnwidth]{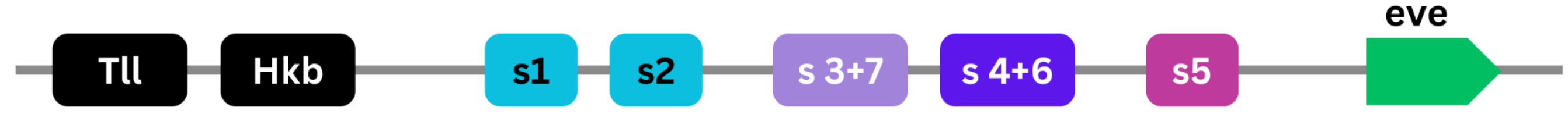}
    \caption{Schematic of the wild-type \textit{eve} locus with two dual stripe elements 3+7 and 4+6 and three single stripe enhancers 1, 2 and 5. Adapted from \cite{Lim2018}. Two repressor locus sites for Tailless (\textit{Tll}) and  Huckebein (\textit{Hkb}) were added, defining the anterior and posterior regions of the \textit{Drsophila}'s embryo where \textit{Eve} expression is absent at steady state.}
    \label{fig:enhancers}
\end{figure}

We tested models with different combinations of repressive and activating transcription factors, but those originating the observed experimental distribution of \textit{Eve} are presented in Figure~\ref{fig:enhancers}. The regulation of the \textit{Eve} stripes as depicted in Figure~\ref{fig:enhancers} is obtained by several combinations of the basic mechanisms \eqref{free-prod}-\eqref{repression}. The choice of wild-type \textit{eve} locus has been adapted from \cite{Lim2018}. We have added two more repressor locus sites for \textit{Tll} and \textit{Hkb}, defining the anterior and posterior regions of the \textit{Drosophila}'s embryo (\cite{Wei} and \cite{Cas}).

The complete list of kinetic mechanisms specific to each enhancer locus for \textit{eve} and \textit{ftz} is presented in the Appendices.  We applied the mass-action law  to each kinetic mechanism and obtained the differential equations describing the expression of the corresponding \textit{Eve} and \textit{Ftz} stripes along the anteroposterior axis of the \textit{Drosophila}'s embryo. The differential equation models to be calibrated were derived using the software package 
\textit{Kinetics} (\cite{kinetics}) written for \textit{Mathematica} programming language.  

We have not considered diffusion mechanisms for proteins or mRNA in the present paper modelling approach. Reaction-diffusion-type models are used to describe the formation of maternal and gap gene families of proteins (\cite{aldi2006}, \cite{DiMu}, \cite{Di14}). Here, we have assumed that the stripes of the pair-rule proteins are formed due to the action of repressive transcription factors distributed along the anteroposterior axis of the \textit{Drosophila}'s embryo, with a clear steady-state expression during the fourteenth mitotic cycle of the \textit{Drosphila} zygote.  

\begin{figure*}[t!]
    \centering
    \includegraphics[width=\textwidth]{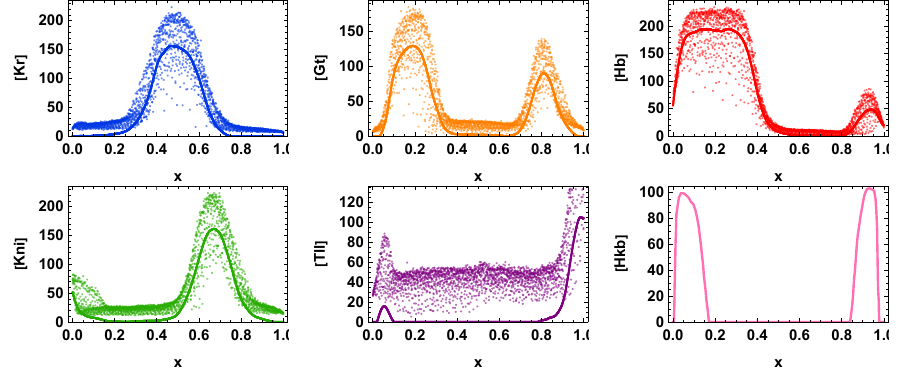}
    \caption{\small{Distribution of gap protein (experimental -- dots -- and filtered) along the anteroposterior axis of the \textit{Drosophila}'s embryo from cycle 14A4.  Ordinate coordinates for each protein are arbitrary and do not correlate. FlyEx datasets: pe3 (\textit{Kr}, \textit{Gt}), rb9 (\textit{Hb}, \textit{Kni}) and th4 (\textit{Tll}). Data for \textit{Hkb} is from the SuperFly database: batch 220310, embryo 016. The embryo lengths were rescaled to $L=1$.}}
    \label{fig:gap_dist}
\end{figure*}

As the distributions of the concentrations of the different proteins along the \textit{Drosophila}'s embryo are not measured in well-defined units, the model equations were adapted so that the variables of interest in the model incorporate a calibration parameter $\alpha_P$, specific to each protein P.  For each protein P, we introduce a new variable 
\begin{equation} \label{alpha}
\overline{\text{P}} = \alpha_{\text{p}} \text{P}
\end{equation}
where $\alpha_{\text{p}}$ is the calibration parameter of a constituent in each reaction. 
Introducing the calibration parameter $\alpha_P$ is necessary because protein concentrations are measured in arbitrary units and differ for the plethora of experiments.
Then, the differential equations in time were solved numerically, with each function taking the experimental spatial distribution of the modelled proteins as input, symbolised by $\overline{\text{P}}$. Model equations with the respective calibration parameters and remaining parameters are listed in the Appendices.

To determine the set of parameters that produce a calibrated steady-state solution that best explains the spatial distribution of \textit{Eve}, we implemented a Monte Carlo algorithm where, at each iteration, every calibration parameter   $\alpha_i$, reaction rate $k_i$ and $d$ were generated randomly in the spatial strip domains of the zygote. A fitness function evaluates the quality of the steady-state solutions of the model equations when compared with the data. 
We represent by $B_{exp}(x_i)$, with $i=1,\ldots, n$, the distribution of the experimental data points along the embryo, where $x_1=0$ and $x_{n}=L=1$ are the anterior and posterior positions of the embryo, respectively. The resulting steady-state distribution of a given protein derived from the model is represented by $B(x_i;\vec{p})$, where $\vec{p} = (p_1, ..., p_m)$ is the set of $m$ simulation parameters. The fitness function is defined as the average of the squared residuals 
$$
    \chi^2(\vec{p}) = \frac{1}{n} \sum_{i=1}^{n} \left(B(x_i;\vec{p}) - B_{exp}(x_i)\right)^2 ,
$$
and the quality of the fit is measured by $\sqrt{\chi^2(\vec{p})}$. The relative error of the fit of stripe $i$ fit is estimated by $\varepsilon_i = \sqrt{\chi^2_i/B^2(x_i;\vec{p})}$, where $x_i$ is the position where the stripe $i$ is maximum  (Materials and methods in  \cite{DiMu2}).

After $M$ iterations of the Monte Carlo algorithm, the set of parameters providing the lower $\chi^2$ value was chosen and used to deliver the best fit to the experimental data of each model.

All model results correspond to the best fit out of half a million trials with Monte Carlo optimisation ($M=0.5\times 10^6$).

Data was taken from the FlyEx database (\cite{Kozlov}, \cite{Myasnikova}, \cite{Poustelnikova}, \cite{Pisarev}), except for \textit{Hkb}, which was taken from the SuperFly database (\cite{SuperFly}). Following common data processing practices (\cite{SURKOVA2008844}), we used a Laplacian filter to remove background noise and levelled all distributions to eliminate residual fluorescent concentrations (\cite{AP}). The size of the embryos was rescaled to L=1. The final gap protein profiles along the normalised anteroposterior position in the embryo of \textit{Drosophila} and the corresponding data are shown in Figure \ref{fig:gap_dist}.

\begin{figure}[H]
    \centering
    \includegraphics[width=0.45\textwidth]{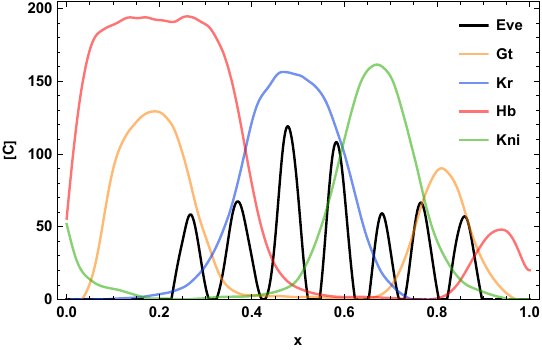}
    \caption{\small{Distribution of \textit{Eve}'s stripes and gap proteins along the \textit{Drosphila}'s embryo. The gap data are from stage 14A4, and \textit{Eve's} data is from stage 14A8. Ordinate coordinates for each protein are arbitrary and do not correlate.}}
    \label{fig:gap_factors}
\end{figure}

To obtain the transcriptional regulators for \textit{Eve}'s stripes, we analysed the distribution of gap proteins along \textit{Drosphila}'s embryo (Figure \ref{fig:gap_factors}).
The observed distributions suggested that the gap transcription factors act as repressors of \textit{eve} and work together combinatorially. 
Therefore, we adopted the concept of individual \textit{eve} enhancers, which only allow \textit{Eve}'s production when none of its repressive regulators binds to it, as expressed by mechanism \eqref{repression}. Assuming a kinetic system where the protein \textit{Eve} is freely produced -- mechanism \eqref{free-prod} with a gene  $\text{G} \to \text{G}_e$ -- and has a subset of repressive gap protein regulators R, each repressor contributes to \textit{Eve's} production and its anteroposterior differentiated distribution. Higher production of \textit{Eve} will occur in regions with lower sums of repressors' concentration, resulting in a localised stripe formation. We tested all the possible combinations of two, three, and four gap proteins, assigning each protein a calibration parameter factor to mimic the calibration parameter $\alpha_P$, and obtained minimal sets of repressive transcriptional regulators. 

In the modelling, we assumed that the concentration of \textit{Eve's} gene, $\overline{\text{G}}_e$ is constant, as mitosis does not occur at the stage of embryogenesis considered, and that $\overline{\text{G}}_e$ has a fixed availability along the embryo. Therefore, we have arbitrarily set  $\overline{\text{G}}_e(0) = 250$ and $\alpha_{\text{G}_e} = 1$. To the reference protein \textit{Eve}, has been assigned the calibration parameter $\alpha_{\text{E}} = 1$, against which each regulator's contribution was measured. At the initial stage of the simulations to calibrate the parameters of the models, all repressor-gene complexes G$_\text{R}$(0) along the zygote were set to zero since they should not be formed at this time. For the other initial conditions, we used the distributions of the corresponding gap regulators in cycle 14A4 shown in Figure \ref{fig:gap_factors}.

\section{Results and Discussion}
\label{sec:resul1}

In the \textit{Drosophila} wild-type embryo, the pair-rule stripes are limited to a specific region of the \textit{Drosophila} embryo, and \textit{Eve} is not present at the anterior and posterior poles. We considered that the \textit{Eve} stripe region is well-determined by the proteins (\textit{Tll}) and (\textit{Hkb}), only produced at the embryo ends. These proteins restrict the expression of \textit{Eve} to the desired region by repressing its transcription at the poles. Therefore, the identified terminal regulators integrated every model developed to describe the formation of \textit{Eve's} stripes.   Moreover, we considered only transcription factors of the gap family of proteins as there is evidence that maternal transcription factors at cell cycle 14 have no transcript activity (\cite{Win}).

\subsection{\textbf{\textit{Eve}'s stripes definition via a gap repressive system}} \label{stripes_model}

\begin{table}[H] 
	\centering
	\resizebox{\columnwidth}{!}{
		\begin{tabular}{c c c c c c c}
		\toprule
		$\mathbf{k_{eve-s1}}$ & $\mathbf{k_{eve-s2}}$ & $\mathbf{k_{eve-s3}}$ & $\mathbf{k_{eve-s4}}$ & $\mathbf{k_{eve-s5}}$ & $\mathbf{k_{eve-s6}}$ & $\mathbf{k_{eve-s7}}$ 
		\\
		\midrule
		0.8584 & 0.8584 & 0.9500 & 3.1100 & 0.9150 & 0.8584 & 1.2900
		\\
		\bottomrule 
	\end{tabular}}
	\caption{\small{\textit{Eve} production rate values for stripe 1-7 that best fit the experimental data for \textit{Eve} at time class 8, cycle 14 of \textit{Drosophila} embryogenesis shown  in Figure \ref{fig:all_eve_fits}.}}
	\label{tab:k1_eve}
\end{table}

\begin{table}[t!] 
	\centering
	\resizebox{\columnwidth}{!}{
		\begin{tabular}{c c c c c c c c}
		\toprule
		& $\mathbf{k_{kni}}$ & $\mathbf{k_{-kni}}$ & $\mathbf{\alpha_{kni}}$ & $\mathbf{k_{hb}}$ & $\mathbf{k_{-hb}}$ & $\mathbf{\alpha_{hb}}$ & d
		\\
		\midrule 
	\textbf{1} 
        & 2.8995 & 0.0002 & 0.5667 & -& - & - & 0.7906\\
        \textbf{2}
        & 5.8021 & 0.0304 & 0.9269 & - & - & - & 0.7906\\
        \textbf{3+7} 
		&  4.0241 & 0.1620 & 0.1153 & 1.4049 & 0.0078 & 0.1523 & 0.7906\\ 
        \textbf{4+6} 
		& 7.4764 & 0.0647 & 0.6720 & 7.1382	 & 0.0424 & 0.0200 & 0.7906\\
        \textbf{5}
        & -	& - & - & 3.6212 & 0.0657 & 0.0103 & 0.7906	\\
		\bottomrule 
		\toprule
		& $\mathbf{k_{kr}}$ & $\mathbf{k_{-kr}}$ & $\mathbf{\alpha_{kr}}$ & $\mathbf{k_{gt}}$ & $\mathbf{k_{-gt}}$ & $\mathbf{\alpha_{gt}}$ &
		\\
		\midrule 
        \textbf{1} 
        & 3.3110 & 0.0001 & 0.0065 & 0.1294 & 0.0098 & 0.5990 & \\
        \textbf{2}
        & 0.7502	& 0.0048 & 0.0419 & 0.1091 & 0.0705 & 0.0819 & 	\\
	\textbf{3+7} 
		& -	& - & - & - & -	& - &  \\ 
        \textbf{4+6} 
		& - & -	& - & 1.5716 & 0.0428 & 0.7840 &  \\
        \textbf{5}
        & 7.4532	& 0.0356 & 0.5312 & 6.5580 & 0.0374 & 0.0894 & 	\\
		\bottomrule	
	\end{tabular}}
	
	\resizebox{\columnwidth}{!}{
		\begin{tabular}{c c c c c c c}
		\toprule 
		$\mathbf{\sqrt{\chi^2_1}}$ & $\mathbf{\sqrt{\chi^2_2}}$ & $\mathbf{\sqrt{\chi^2_3}}$ & $\mathbf{\sqrt{\chi^2_4}}$ & $\mathbf{\sqrt{\chi^2_5}}$ & $\mathbf{\sqrt{\chi^2_6}}$ & $\mathbf{\sqrt{\chi^2_7}}$ \\
		\midrule
		4.1 & 3.2 & 13.2 & 36.1 & 4.3 & 17.8 & 19.6\\
		\bottomrule 
		\toprule
		$\mathbf{\varepsilon_1}$ & $\mathbf{\varepsilon_2}$ & $\mathbf{\varepsilon_3}$ & $\mathbf{\varepsilon_4}$ & $\mathbf{\varepsilon_5}$ & $\mathbf{\varepsilon_6}$ & $\mathbf{\varepsilon_7}$ \\
		\midrule
		0.07 & 0.05 & 0.11 & 0.33 & 0.07 & 0.27 & 0.33 \\ 
		\bottomrule
	\end{tabular}}
	\caption{\small{Parameter values that best fit the experimental data of \textit{Eve's} stripes at time class 8 of mitotic cycle 14 of \textit{Drosophila} embryogenesis shown in Figure \ref{fig:all_eve_fits}. Terminal gap repression parameters were fixed with the 3+7 simulation: $k_{tll} = 2.2956$, $k_{-tll} = 0.0747$, $\alpha_{tll} = 0.7925$, $k_{hkb} = 2.4229$, $k_{-hkb} = 10^{-5}$, $\alpha_{hkb} = 0.7988$. Model results after 500,000 trials with Monte Carlo, with calibration parameters $\alpha$ determined within the interval [0,1]. The quantity $\sqrt{\chi^2(\vec{p})}=\sqrt{\chi^2_i}$ measures the quality of the fit of stripe number $i$, and $\varepsilon_i = \sqrt{\chi^2_i/B^2(x_i;\vec{p})}$ is an estimate of the relative error of each stripe fit.}}
	\label{tab:fits_eve}
\end{table}

In the first attempt of the model calibration, we considered that the production rate of \textit{Eve} was the same across all cells. However, we found that adjusting \textit{Eve}'s production spatially was necessary to improve the accuracy of the fits. For this reason, we considered the production rate as a function of space and adjusted it stripe by stripe to best match the data (Table \ref{tab:k1_eve}).   This approach is supported by recent studies demonstrating mRNA translation's profound heterogeneity, including protein synthesis rates (\cite{Sonneveld2020}).

\subsubsection{Single stripe 1 enhancer of \textit{Eve}}

To explain the formation of stripe 1 of the \textit{Eve} protein, we propose a novel enhancer model regulated by the repressive proteins \textit{Tll}, \textit{Hkb}, \textit{Kr}, \textit{Kni} and \textit{Gt}, with kinetic mechanisms
\begin{equation} \label{S1-2system}
    \begin{array}{ll}
        \displaystyle
        \text{G}_e \mathop{\longrightarrow}^{k_{1}} \text{G}_e + \text{Eve}, & \displaystyle
        \text{Eve} \mathop{\longrightarrow}^{d}, \\ \displaystyle
        \text{Kr} + \text{G}_e \mathop{\rightleftarrows}^{k_{3}}_{k_{-3}} \text{G}_{R1}, & \displaystyle
        \text{Kni} + \text{G}_e \mathop{\rightleftarrows}^{k_{4}}_{k_{-4}} \text{G}_{R2},\\ \displaystyle
        \text{Gt} + \text{G}_e \mathop{\rightleftarrows}^{k_{5}}_{k_{-5}} \text{G}_{R3}, & \displaystyle
        \text{Tll} + \text{G}_e \mathop{\rightleftarrows}^{k_{6}}_{k_{-6}} \text{G}_{R4}, \\ \displaystyle
        \text{Hkb} + \text{G}_e \mathop{\rightleftarrows}^{k_{7}}_{k_{-7}} \text{G}_{R5}, &
    \end{array}
\end{equation} 
where $\text{G}_e$ is the gene of  \textit{Eve}.
The corresponding evolution equations derived from the mass action law are shown in \ref{AppA}. The parameters of the model that best fit the data for stripe 1 are shown in Tables~\ref{tab:k1_eve} and \ref{tab:fits_eve}. 
The parameters in the kinetic mechanisms \eqref{S1-2system} in Tables~\ref{tab:k1_eve} and \ref{tab:fits_eve} translate to: (Table~\ref{tab:k1_eve}) $k_1=k_{eve-s1}$, (Table~\ref{tab:fits_eve}) $k_3=k_{kr}$, $k_{-3}=k_{-kr}$, $k_4= k_{kni}$, $k_{-4}=k_{-kni}$, $k_5=k_{gt}$, $k_{-5}=k_{-gt}$, $k_6=k_{tll}$, $k_{-6}=k_{-tll}$, $k_7=k_{hkb}$ and $k_{-7}=k_{-hkb}$.

In Figure \ref{fig:all_eve_fits}, we show the position and shape of stripe 1 along the anteroposterior axis of the \textit{Drosophila} embryo, showing a remarkable agreement with the experimental data, reflected in a relative error of   $\varepsilon_1=7\%$ (Table~\ref{tab:fits_eve}).

\subsubsection{Single stripe 2 enhancer of \textit{Eve}}

When testing the joint formation of stripes 2 and 5 of \textit{Eve} with only \textit{Gt} and \textit{Kr}, we realised that the only option was to separate their regulation with two single stripe enhancers. Thus, we propose \textit{Kr}, \textit{Kni}, and \textit{Gt} to be the minimal gap set required to explain stripe 2, just as for stripe 1, with the same kinetic mechanisms \eqref{S1-2system}, but different parameters. In this case, (Table~\ref{tab:k1_eve}) $k_1=k_{eve-s2}$ and the other parameters are defined as in the stripe 1 case.

With this assumption, we obtained a very precise prediction of stripe 2 -- Figure \ref{fig:all_eve_fits}, with a relative error lower than $\varepsilon_2=5\%$ (Table~\ref{tab:fits_eve}). Whereas for stripe 1 \textit{Gt} and \textit{Kni} had more or less the same level of repression, now we find a greater affinity to both \textit{Gt} and \textit{Kr} compared to \textit{Kni} (Table \ref{tab:fits_eve}).

\subsubsection{Dual stripe enhancer 3+7 of \textit{Eve}}

We propose a dual stripe enhancer model for stripes 3 and 7 of \textit{Eve} with the repressive regulators \textit{Tll}, \textit{Hkb}, \textit{Kni} and \textit{Hb}:
\begin{equation} \label{S37system}
    \begin{array}{ll}
        \displaystyle
        \text{G}_e \mathop{\longrightarrow}^{k_{1}} \text{G}_e + \text{Eve}, & \displaystyle
        \text{Eve} \mathop{\longrightarrow}^{d} , \\\displaystyle 
        \text{Kni} + \text{G}_e \mathop{\rightleftarrows}^{k_{3}}_{k_{-3}} \text{G}_{R1}, & \displaystyle
       \text{Hb} + \text{G}_e \mathop{\rightleftarrows}^{k_{4}}_{k_{-4}} \text{G}_{R2},\\\displaystyle 
        \text{Tll} + \text{G}_e \mathop{\rightleftarrows}^{k_{5}}_{k_{-5}} \text{G}_{R3}, & \displaystyle
        \text{Hkb} + \text{G}_e \mathop{\rightleftarrows}^{k_{6}}_{k_{-6}} \text{G}_{R4}.
    \end{array}
\end{equation}
The equations associated with these kinetic mechanisms are shown in \ref{AppA}.

The parameters best fitting the data for stripes 3 and 7 are shown in Tables~\ref{tab:k1_eve} and \ref{tab:fits_eve}. 
The parameters in the kinetic mechanisms \eqref{S37system} correspond to the following parameters in Tables~\ref{tab:k1_eve} and \ref{tab:fits_eve}: (Table~\ref{tab:k1_eve}) $k_1=k_{eve-s3}$, or $k_1=k_{eve-s7}$, or $k_1=0.8584$ away from the stripes 3 and 7 regions; (Table~\ref{tab:fits_eve}) $k_3=k_{kni}$, $k_{-3}=k_{-kni}$, $k_4=k_{hb}$, $k_{-4}=k_{-hb}$, $k_5=k_{tll}$, $k_{-5}=k_{-tll}$, $k_6=k_{hkb}$ and $k_{-6}=k_{-hkb}$.

\begin{figure*}[t!]
    \centering
    \includegraphics[width=\textwidth]{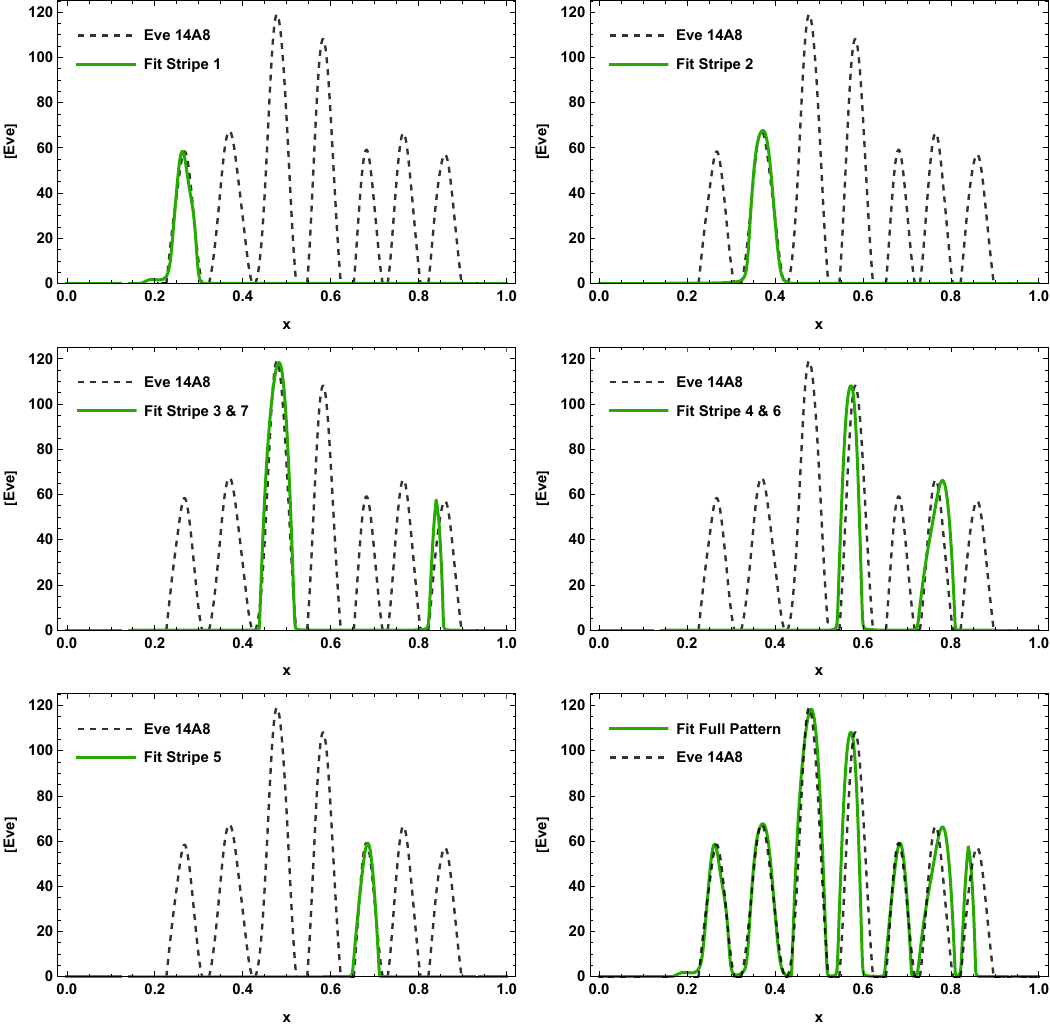}
    \caption{\small{Predictions for \textit{eve's} stripe-specific enhancers 1, 2, 5, 3+7 and 4+6 and full \textit{Eve} pattern prediction obtained by combining the results of each enhancer model. The figure shows the best fits of each model to the experimental data of \textit{Eve} stripes  (dotted lines) at time class 8 of mitotic cycle 14 (14A8) of \textit{Drosophila} embryogenesis. We obtained the parameter values of k$_i$ and $\alpha_i$ by a Monte Carlo optimisation after 500,000 trials, summarised in Table \ref{tab:fits_eve}}. The adjusted \textit{Eve} production rates k$_1$ are shown in Table \ref{tab:k1_eve}.}
    \label{fig:all_eve_fits}
\end{figure*}

The steady-state solution predicted -- Figure \ref{fig:all_eve_fits} -- shows the location, width, and amplitude of stripe 3 practically match the data, reflected in a low relative error of   $\varepsilon_3=11\%$. Concerning stripe 7, we see that the anterior border forms correctly, but its location and width are not precise, with a high relative error of  $\varepsilon_7=33\%$. This may be due to an excessive \textit{Hkb} repression in the posterior embryo pole. The \textit{Hkb} profile overlaps with stripe 7, suggesting that our model will always underestimate this stripe with the initial conditions data we had access to. 

We showed that the proper functioning of this enhancer requires \textit{Kni} to have a slightly higher repression calibration parameter, and at a faster rate, than \textit{Hb} (Table \ref{tab:fits_eve}).

\subsubsection{Dual stripe enhancer 4+6 of \textit{Eve}}

For stripes 4 and 6 of the \textit{Eve} pattern, we propose a model with the repressive regulators \textit{Tll}, \textit{Hkb}, \textit{Kni} and \textit{Gt} and \textit{Hb}:
\begin{equation}  \label{S46system}
 \begin{array}{ll}
        \displaystyle
        \text{G}_e \mathop{\longrightarrow}^{k_{1}} \text{G}_e + \text{Eve}, & \displaystyle
        \text{Eve} \mathop{\longrightarrow}^{d}, \\
        \text{Kni} + \text{G}_e \mathop{\rightleftarrows}^{k_{3}}_{k_{-3}} \text{G}_{R1}, & \displaystyle
        \text{Gt} + \text{G}_e \mathop{\rightleftarrows}^{k_{4}}_{k_{-4}} \text{G}_{R2}, \\ \displaystyle
        \text{Hb} + \text{G}_e \mathop{\rightleftarrows}^{k_{5}}_{k_{-5}} \text{G}_{R3},& \displaystyle
        \text{Tll} + \text{G}_e \mathop{\rightleftarrows}^{k_{6}}_{k_{-6}} \text{G}_{R4}, \\ \displaystyle
        \text{Hkb} + \text{G}_e \mathop{\rightleftarrows}^{k_{7}}_{k_{-7}} \text{G}_{R5}. &
    \end{array}
\end{equation}
The equations associated with these kinetic mechanisms are shown in \ref{AppA}. The parameters best fitting the data for stripes 4 and 6 are shown in Tables~\ref{tab:k1_eve} and \ref{tab:fits_eve}. 
The parameters in the kinetic mechanisms \eqref{S46system} correspond to the following parameters in Tables~\ref{tab:k1_eve} and \ref{tab:fits_eve}: (Table~\ref{tab:k1_eve}) $k_1=k_{eve-s4}$, or $k_1=k_{eve-s6}$, or $k_1=0.8584$ away from the stripes 4 and 6 regions; (Table~\ref{tab:fits_eve}) $k_3=k_{kni}$, $k_{-3}=k_{-kni}$, $k_4=k_{gt}$, $k_{-4}=k_{-gt}$, $k_5=k_{hb}$, $k_{-5}=k_{-hb}$, $k_6=k_{tll}$, $k_{-6}=k_{-tll}$, $k_7=k_{hkb}$ and $k_{-7}=k_{-hkb}$.

Regarding stripes 4 and 6, we achieved an overall correct prediction of both stripe's localisation (Figure \ref{fig:all_eve_fits}). However, localisation deviations of both stripes significantly increase their relative errors to around $\varepsilon_4=33\%$ and $\varepsilon_6=27\%$.

In this case, we found it necessary for \textit{Hb} to have a much higher repressive contribution than \textit{Kni} and \textit{Gt}, contrary to the 3+7 case. This sensitivity difference was proposed in the literature (\cite{Clyde}, \cite{Crocker2016QuantitativelyPC}) and is now confirmed with our simple enhancer models. Furthermore, our results prove to be consistent with the computationally predicted number of binding sites for each enhancer: \cite{Clyde} identify 12 \textit{Hb} and 16 \textit{Kni} binding sites in the 3+7 enhancer, and 11 \textit{Hb} and 4 \textit{Kni} binding sites in the 4+6 enhancer. The proportions between these numbers are very similar to those between the calibration $\alpha's$ we computed: $\alpha_{Kni} = 0.1152$ and $\alpha_{Hb} = 0.1523$ for the 3+7 enhancer, and $\alpha_{Kni} = 0.6720$ and $\alpha_{Hb} = 0.0200$ for the 4+6 enhancer. These differences may be associated with the quality of different data sets.

\subsubsection{Single stripe 5 enhancer of \textit{Eve}}

We found that \textit{Tll}, \textit{Hkb}, \textit{Kr}, and \textit{Gt} are sufficient to generate stripe 5 of \textit{Eve}. However, when we added \textit{Hb} as a repressive regulator, we improved the fit to the data, lowering the relative error from around $\varepsilon_5\simeq13\%$  to less than $\varepsilon_5=8\%$  (Figure \ref{fig:all_eve_fits}, Table \ref{tab:fits_eve}). This latter enhancer has \textit{Hb} as the main repressive regulator, followed by \textit{Gt}, and both have much higher calibration parameter factors than \textit{Kr}. Therefore, to regulate stripe 5, we have chosen the kinetic mechanism
\begin{equation}  \label{S5system2}
    \begin{array}{ll}
        \displaystyle
        \text{G}_e \mathop{\longrightarrow}^{k_{1}} \text{G}_e + \text{Eve}, & \displaystyle
        \text{Eve} \mathop{\longrightarrow}^{d}, \\ \displaystyle
        \text{Kr} + \text{G}_e \mathop{\rightleftarrows}^{k_{3}}_{k_{-3}} \text{G}_{R1} ,& \displaystyle
        \text{Gt} + \text{G}_e \mathop{\rightleftarrows}^{k_{4}}_{k_{-4}} \text{G}_{R3}, \\ \displaystyle
        \text{Hb} + \text{G}_e \mathop{\rightleftarrows}^{k_{5}}_{k_{-5}} \text{G}_{R4}, & \displaystyle
        \text{Tll} + \text{G}_e \mathop{\rightleftarrows}^{k_{6}}_{k_{-6}} \text{G}_{R5},\\ \displaystyle
        \text{Hkb} + \text{G}_e \mathop{\rightleftarrows}^{k_{7}}_{k_{-7}} \text{G}_{R6} .&
    \end{array}
\end{equation}
The equations associated with these kinetic mechanisms are shown in \ref{AppA}. The parameters that best fit the data for stripe 5 are shown in Tables~\ref{tab:k1_eve} and \ref{tab:fits_eve}. 
The parameters in the kinetic mechanisms \eqref{S5system2} correspond to the following parameters in Tables~\ref{tab:k1_eve} and \ref{tab:fits_eve}: (Table~\ref{tab:k1_eve}) $k_1=k_{eve-s5}$, (Table~\ref{tab:fits_eve}) $k_3=k_{kr}$, $k_{-3}=k_{-kr}$, $k_4=k_{gt}$, $k_{-4}=k_{-gt}$, $k_5=k_{hb}$, $k_{-5}=k_{-hb}$, $k_6=k_{tll}$, $k_{-6}=k_{-tll}$, $k_7=k_{hkb}$ and $k_{-7}=k_{-hkb}$.

Overall, we accurately predicted all the stripes in the \textit{Eve} pattern with a purely repressive gap regulating system. We found no justification or need for any gap transcription activator role. Moreover, our model does not require any maternal input. We propose that any effect on the \textit{Eve} pattern observed in maternal gene mutation experiments comes from collateral damage to gap regulation and consequent disturbance in the formation of the gap domains.

\subsection{\textbf{Simulating gap mutations of \textit{Eve}}}

\begin{figure*}[t]
    \centering
    \includegraphics[width=0.99\textwidth]{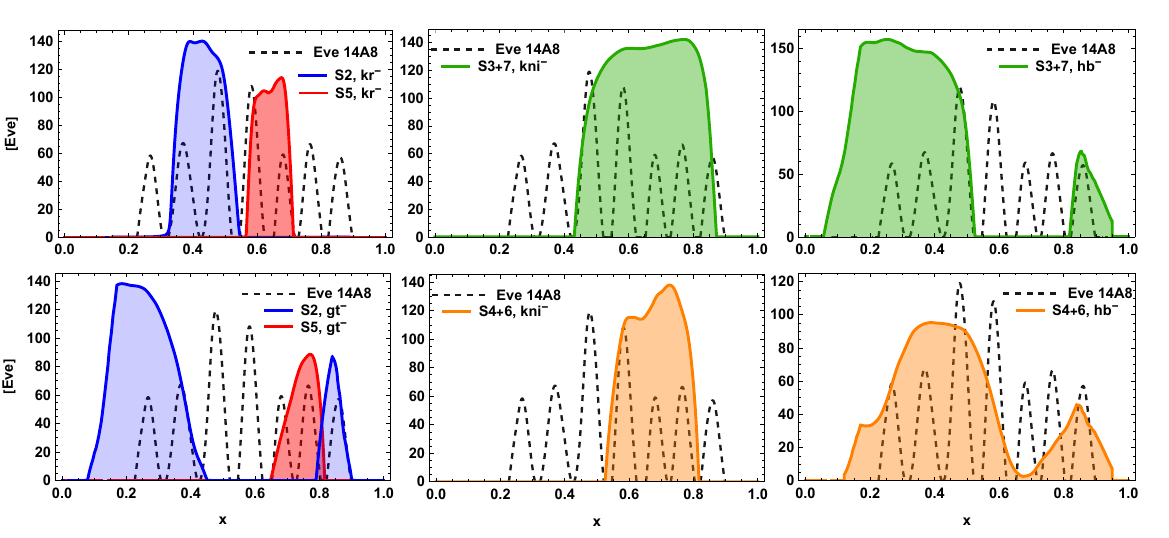}
    \caption{\small{Effects on \textit{Eve's} pattern of the introduction of gap gene mutations $kr^-$, $gt^-$, $kni^-$ and $hb^-$ in each enhancer model, cycle 14A8 of \textit{Drosophila} embryogenesis. A $kr^-$ embryo shows two fused domains, 2-3 and 4-5, coincident with the experimental results of \cite{SURKOVA2008844} and \cite{Fuji}. A $gt^-$ embryo presents a merging of stripes 1 and 2 as of stripes 5 and 6, in agreement with \cite{Frash_correct}, \cite{Small1991} and \cite{Stano}. A $kni^-$ embryo presents a merging of stripes in the domain 3 to 7, just as observed in \cite{SURKOVA2008844}. A $hb^-$ embryo shows an anterior fused domain up to stripe 3 or 4, a posterior expansion of stripe 7, and a possible overlap with stripe 6. This is in line with the mutation experiments studied in \cite{Frash_correct}, \cite{Small1991} and \cite{Fuji}.}}
    \label{fig:mutations}
\end{figure*}

An essential step in validating our model was the introduction of gap mutations. We removed one regulator at a time in each model and observed the effects on the \textit{Eve} pattern of time class 8 of mitotic cycle 14 (Figure \ref{fig:mutations}).

For a \textit{kr} mutation introduced in stripe 2 (Figure \ref{fig:mutations}, S2, \textit{kr}$^-$) and stripe 5 enhancers (Figure \ref{fig:mutations}, S5, \textit{kr}$^-$), we observe the fusion of stripes 2 and 3 and the fusion of stripes 4 and 5. Such a result coincides with the experimental results of \cite{Surkova2013} and with the anterior expansion of stripe 5 reported in \cite{Fuji}. However, we do not observe the weakening of stripe 4 and the posterior expansion of stripe 6, reported in that same article, or even a more pronounced expression of stripe 2. We point out that we are not considering the collateral effects of each mutation in the expression of other gap genes. Thus, we disregard any indirect alteration of \textit{Eve's} pattern. For instance, a \textit{kr} mutation can influence all other gap genes \textit{gt}, \textit{kni} and \textit{hb}  involved in forming stripes 4 and 6 (\cite{aldi2006}). 

Next, we analysed the introduction of a \textit{gt} mutation (Figure \ref{fig:mutations}, S2 and S5, \textit{gt}$^-$), in which stripes 1 and 2 merge as well as stripes 5 and 6. These results agree with the experiments carried out by \cite{Frash_correct} and also with \cite{Small1991} and \cite{Stano} that report the expansion of the anterior border of stripe 2. Although an intensity reduction for stripe 7 is shown in \cite{Frash_correct}, we do not verify that. On the contrary, we observe the appearance of a seventh stripe that did not exist before on this model.

As for the \textit{kni} mutation, it was introduced in both enhancers 3+7 and 4+6 (Figure \ref{fig:mutations}, S3+7, \textit{kni}$^-$ and S4+6, \textit{kni}$^-$). They show a fusion of stripes 3 to 7 and a fusion of stripes 4 to 6, the behaviour expected according to \cite{Surkova2013} experiments.

Finally, we studied the introduction of an \textit{hb} mutation in the 3+7 and 4+6 enhancers (Figure \ref{fig:mutations}, S3+7, \textit{hb}$^-$ and S4+6, \textit{hb}$^-$). For the 3+7 case, we observe a broad fusion domain starting at the anterior end of the embryo and encompassing stripes 1, 2 and 3 and a posterior expansion of stripe 7. As for the 4+6 enhancer, we see a fused expression domain in the anterior region up to stripe 4 and an expansion of the posterior border of stripe 6, overlapping with stripe 7. Both cases are in line with the mutation experiences made by \cite{Frash_correct}, \cite{SMALL1996314} and \cite{Fuji} as they report a total or partial deletion of stripes 1 to 4, with a broad anterior band of expression instead, the expansion of stripes 3 and 7, and the expansion of stripes 4 and 6.

Both \textit{tll} and \textit{hkb} mutations were tested, but they did not cause significant changes to \textit{Eve's} pattern aside from a lack of definition in the embryo poles. Contrarily to what was expected (\cite{SMALL1996314}, \cite{Frash_correct}, \cite{Fuji}, \cite{Janssens2013}, \cite{Ilsley2013}), a \textit{tll} mutation did not abolish stripe 7. Once more, we believe that the abolition of this stripe is somehow related to the direct interference with the regulation of other gap genes, such as \textit{kni} and \textit{gt}, that is caused by a \textit{tll} mutation. Furthermore, we hypothesise that the experimentally recorded effects of a mutation in maternal genes \textit{nanos} (\textit{nos}) and \textit{torso} (\textit{tor}) are once again the effect of the mis-expression of gap genes. For example, \textit{tor} regulates the terminal gap genes, so a \textit{tor} mutation is expected to alter \textit{Eve} in a similar way to a mutation in \textit{tll} or \textit{hkb}.

\subsection{\textbf{\textit{Ftz}'s stripes can be partially defined via a gap repressive system}} \label{stripes_model_ftz}

\begin{figure}[h]
    \centering
    \includegraphics[width=\columnwidth]{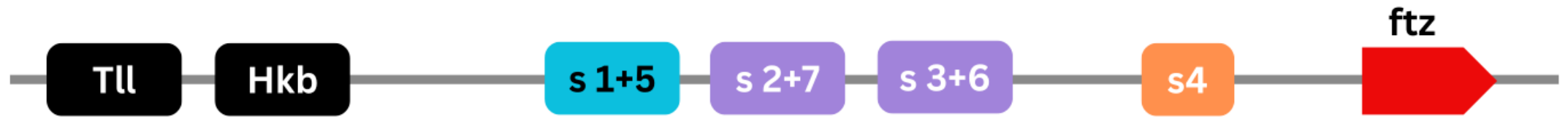}
    \caption{Schematic of the hypothesised regulation of the wild-type \textit{ftz} locus with three dual stripe elements 1+5, 2+7 and 3+6 stripe enhancers. Enhancers for stripe 4 are unknown.}
    \label{fig:figftz}
\end{figure}

Following a reasoning similar to the formation of \textit{Eve} stripes, we tested the viability of stripe-specific enhancers driving \textit{Ftz} stripes. The gap protein combinations we identified as possibly regulating \textit{Ftz} correspond to those found to regulate \textit{eve's} transcription: our thermodynamics approach is compatible with two dual stripe elements 2+7 and 3+6 regulated by \textit{Hb} and \textit{Kni} and another dual stripe enhancer 1+5 regulated by \textit{Kr} and \textit{Gt}. We also assume that \textit{Tll} and \textit{Hkb} repress the production of \textit{Ftz} in the anterior and posterior regions of the \textit{Drosophyla}'s embryo (Figure~\ref{fig:figftz}).

\begin{figure*}[h]
    \centering
    \includegraphics[width=\textwidth]{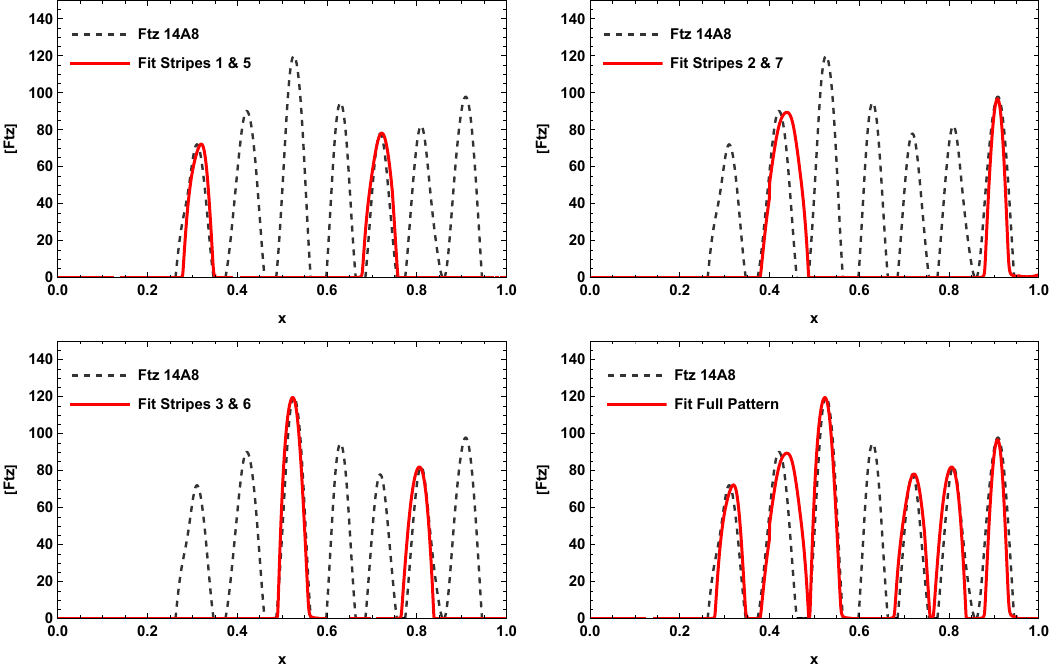}
    \caption{\small{Predictions for \textit{ftz's} stripe-specific enhancers 1+5, 2+7, 3+6 and full \textit{Ftz} pattern prediction obtained by combining the results of each enhancer model. The figure shows the best fits of each model to the experimental data of \textit{Ftz}  stripes (dotted lines) at time class 8 of mitotic cycle 14 (14A8) of \textit{Drosophila}'s embryogenesis. We obtained the parameter values of k$_i$ and $\alpha_i$ by a Monte Carlo optimisation after 500,000 trials, summarised in Table \ref{tab:fits_ftz}}. The adjusted \textit{Ftz} production rates k$_1$ are listed in Table \ref{tab:k1_ftz}.}
    \label{fig:all_ftz_fits}
\end{figure*}

\subsubsection{Dual stripe enhancer 1+5 of \textit{Ftz}}

We initially thought that \textit{Gt} and \textit{Kr} were enough to regulate the 1+5 enhancer, but we discovered that adding a third regulator, \textit{Kni}, is essential. In this case, the kinetic regulatory mechanism is similar to the one for stripes 1 and 2 of \textit{Eve} -- kinetic mechanism \eqref{S1-2system} --  with $\text{G}_e\to \text{G}_f$,
where $\text{G}_f$ is the gene of \textit{Ftz}. The corresponding evolution equations derived from the mass action law are shown in \ref{AppA}. The parameters that best fit the data for stripes 1 and 5 are shown in Tables~\ref{tab:k1_ftz} and \ref{tab:fits_ftz}:
 (Table~\ref{tab:k1_ftz}) $k_1=k_{ftz-s1}$, or $k_1=k_{ftz-s5}$, or {$k_1=0.5823$ away from the stripes 1 and 5 regions}; (Table~\ref{tab:fits_ftz}) $k_3=k_{kr}$, $k_{-3}=k_{-kr}$, $k_4=k_{kni}$, $k_{-4}=k_{-kni}$, $k_5=k_{gt}$, $k_{-5}=k_{-gt}$, $k_6=k_{tll}$ and $k_{-6}=k_{-tll}$, $k_7=k_{hkb}$ and $k_{-7}=k_{-hkb}$.

In Figure \ref{fig:all_ftz_fits}, we show the position of stripes 1 and 5 along the anteroposterior axis of the \textit{Drosophila}'s embryo, with a remarkable agreement with the experimental data, reflected in relative errors of $\varepsilon_1=20\%$ and $\varepsilon_5=18\%$, respectively (Table~\ref{tab:fits_eve}).

We found this enhancer to have a higher affinity to \textit{Kr} and \textit{Kni} than to \textit{Gt} (Table \ref{tab:fits_ftz}).

\begin{table}[h!] 
	\centering
	\resizebox{\columnwidth}{!}{
		\begin{tabular}{c c c c c c c}
		\toprule
		$\mathbf{k_{ftz-s1}}$ & $\mathbf{k_{ftz-s2}}$ & $\mathbf{k_{ftz-s3}}$ & $\mathbf{k_{ftz-s4}}$ & $\mathbf{k_{ftz-s5}}$ & $\mathbf{k_{ftz-s6}}$ & $\mathbf{k_{ftz-s7}}$ 
		\\
		\midrule
		1.3100 & 0.6800 & 1.8910 & - & 1.500 & 0.5630 & 3.3400
		\\
		\bottomrule 
	\end{tabular}}
	\caption{\small{\textit{Ftz} production rate values that best fit the experimental data at time class 8 of mitotic cycle 14 of \textit{Drosophila} embryogenesis shown in Figure \ref{fig:all_ftz_fits}.}}
	\label{tab:k1_ftz}
\end{table}

\begin{table}[t!] 
	\centering
	\resizebox{\columnwidth}{!}{
		\begin{tabular}{c c c c c c c c}
		\toprule
		& $\mathbf{k_{kni}}$ & $\mathbf{k_{-kni}}$ & $\mathbf{\alpha_{kni}}$ & $\mathbf{k_{hb}}$ & $\mathbf{k_{-hb}}$ & $\mathbf{\alpha_{hb}}$ & d
		\\
		\midrule 
         \textbf{1+5} 
 		& 5.6887 & 0.0205 & 0.9909 & - & - & - & 0.7104\\
        \textbf{2+7} 
 		& 7.3807 & 0.0034 & 0.0434 & 4.5349 & 0.0375 & 0.5605 & 0.7104\\ 
        \textbf{3+6} 
        & 3.1028 & 0.0877 & 0.3160 & 7.7090 & 0.0265 & 0.0440 & 0.7104\\
		\bottomrule 
		\toprule
		& $\mathbf{k_{kr}}$ & $\mathbf{k_{-kr}}$ & $\mathbf{\alpha_{kr}}$ & $\mathbf{k_{gt}}$ & $\mathbf{k_{-gt}}$ & $\mathbf{\alpha_{gt}}$ &  
		\\
		\midrule 
		\textbf{1+5} 
 		& 3.6834 & 0.0748 & 0.2489 & 0.3609 & 0.0001 & 0.3461 &  \\
		\textbf{2+7} 
 		& - & - & - & - & - & - & \\ 
        \textbf{3+6} 
        & - & - & - & - & - & - & \\
		\bottomrule	
	\end{tabular}}
	
	\resizebox{\columnwidth}{!}{
		\begin{tabular}{c c c c c c c}
		\toprule 
		$\mathbf{\sqrt{\chi^2_1}}$ & $\mathbf{\sqrt{\chi^2_2}}$ & $\mathbf{\sqrt{\chi^2_3}}$ & $\mathbf{\sqrt{\chi^2_4}}$ & $\mathbf{\sqrt{\chi^2_5}}$ & $\mathbf{\sqrt{\chi^2_6}}$ & $\mathbf{\sqrt{\chi^2_7}}$ \\
		\midrule
		14.2 & 32.4 & 10.3 & - & 14.1 & 15.9 & 18.0\\
		\bottomrule 
		\toprule
		$\mathbf{\varepsilon_1}$ & $\mathbf{\varepsilon_2}$ & $\mathbf{\varepsilon_3}$ & $\mathbf{\varepsilon_4}$ & $\mathbf{\varepsilon_5}$ & $\mathbf{\varepsilon_6}$ & $\mathbf{\varepsilon_7}$ \\
		\midrule
		0.20 & 0.36& 0.09 & - & 0.18 & 0.19 & 0.18 \\ 
		\bottomrule
	\end{tabular}}
	\caption{\small{Parameter values that best fit the experimental data of \textit{Ftz's} stripes at time class 8 of mitotic cycle 14 of \textit{Drosophila}'s embryogenesis shown in Figure \ref{fig:all_eve_fits}. Terminal gap repression parameters fixed with the 2+7 simulation for \textit{Tll} and with the 1+5 simulation for \textit{Hkb}: $k_{tll} = 5.2564$, $k_{-tll} = 0.0937$, $\alpha_{tll} = 0.4560$, $k_{hkb} = 8.3229$, $k_{-hkb} = 0.0500$, $\alpha_{hkb} = 0.2900$. Model results after 500,000 trials with Monte Carlo, with calibration parameters $\alpha$ determined within the interval [0,1]. The quantity $\sqrt{\chi^2(\vec{p})}=\sqrt{\chi^2_i}$ measures the quality of the fit of stripe number $i$, and $\varepsilon_i = \sqrt{\chi^2_i/B^2(x_i;\vec{p})}$ is an estimate of the relative error of each stripe fit.}}
	\label{tab:fits_ftz}
\end{table}

\subsubsection{Dual stripe enhancer 2+7 of \textit{Ftz}}

The position of stripes 2 and 7 are predicted with an enhancer model with repressive regulators \textit{Hb}, \textit{Kni},  \textit{Tll} and \textit{Hkb}. In this case, the kinetic regulatory mechanism is similar to the one for dual stripes 3+7 of \textit{Eve}, \eqref{S37system}, with $\text{G}_e\to \text{G}_f$.
The corresponding evolution equations derived from the mass action law are shown in \ref{AppA}. The parameters that best fit the data for stripes 2 and 7 are shown in Tables~\ref{tab:k1_ftz} and \ref{tab:fits_ftz}:
 (Table~\ref{tab:k1_ftz}) $k_1=k_{ftz-s2}$, or $k_1=k_{ftz-s7}$, or {$k_1=0.5823$ away from the stripes 2 and 7 regions}; (Table~\ref{tab:fits_ftz}) $k_3=k_{kni}$, $k_{-3}=k_{-kni}$, $k_4=k_{hb}$, $k_{-4}=k_{-hb}$, $k_5=k_{tll}$ and $k_{-5}=k_{-tll}$, $k_6=k_{hkb}$ and $k_{-6}=k_{-hkb}$.

An enhancer model with repressive regulators \textit{Hb} and \textit{Kni} predicted stripe 7 with a reasonable agreement with the data and a low relative error of about $\varepsilon_7=18\%$. Its location was precise, but its width was underestimated. As for stripe 2, it has a well-described anterior border but an extended posterior border. Such can be an indication that the \textit{Hb} repression calibration parameter determined ($\alpha_{Hb} = 0.5605$) is not enough. Together with the width mismatch, this leads to a high relative error of $\varepsilon_2=36\%$. 

The proper functioning of this enhancer requires a higher repression calibration parameter from \textit{Kni} versus \textit{Hb} (Table \ref{tab:fits_ftz}).

\subsubsection{Dual stripe enhancer 3+6 of \textit{Ftz}}

Finally, we studied the control mechanisms of the dual stripe enhancer 3+6, which we propose to be regulated by the same proteins as the 2+7 enhancer of \textit{Ftz} and the 3+7 enhancer of \textit{Eve}: \textit{Hb}, \textit{Kni}, \textit{Tll} and \textit{Hkb}. The predicted stripes match the experimental data, reflected in the low relative errors for stripes 3 and 6 of $\varepsilon_3=9\%$ and $\varepsilon_6=19\%$. 

Contrarily to what is observed for the dual stripe enhancer 2+7, 
the correct functioning of the 3+6 enhancer relies on a higher concentration of \textit{Hb} versus \textit{Kni} ($\alpha_{Hb} = 0.0440$ and $\alpha_2 = 0.3160$). Thus, despite depending on the same regulators as the 2+7, enhancer 3+6 has a greater affinity or a greater number of binding sites for \textit{Hb} than for \textit{Kni}. Note that this relationship between the \textit{ftz's} 2+7 and 3+6 enhancers is the same as between \textit{eve's} 3+7 and 4+6 enhancers. If these two dual stripe enhancers are indeed involved in \textit{Ftz}'s pattern formation, it would be interesting to determine experimentally the number of binding sites in each and compare them to the calibration $\alpha$'s we computationally predicted.

We proved that stripe-specific elements can partially define the \textit{Ftz}  pattern along the anteroposterior axis of the \textit{Drosophila}'s embryo, with stripe 4 being the only one that can't be explained with gap transcription factors.

\begin{figure}[H]
    \centering
    \includegraphics[width=0.95\columnwidth]{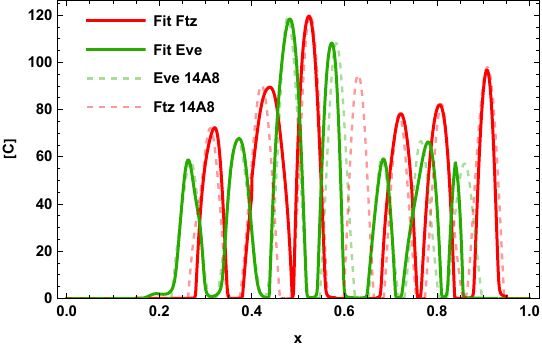}
    \caption{\small{Best fits of the experimental data of the full \textit{Eve} and \textit{Ftz} patterns at time class 8 of mitotic cycle 14 (14A8) of \textit{Drosophila} embryogenesis. These predictions were obtained following a stripe-specific enhancer theory. No regulators were identified for \textit{Ftz's} stripe 4.}}
    \label{fig:full_ftz_eve}
\end{figure}

\subsection{\textbf{Final predictions for \textit{Eve} and \textit{Ftz} patterns}}

 \begin{figure*}[t!]
    \centering
    \includegraphics[width=0.95\textwidth]{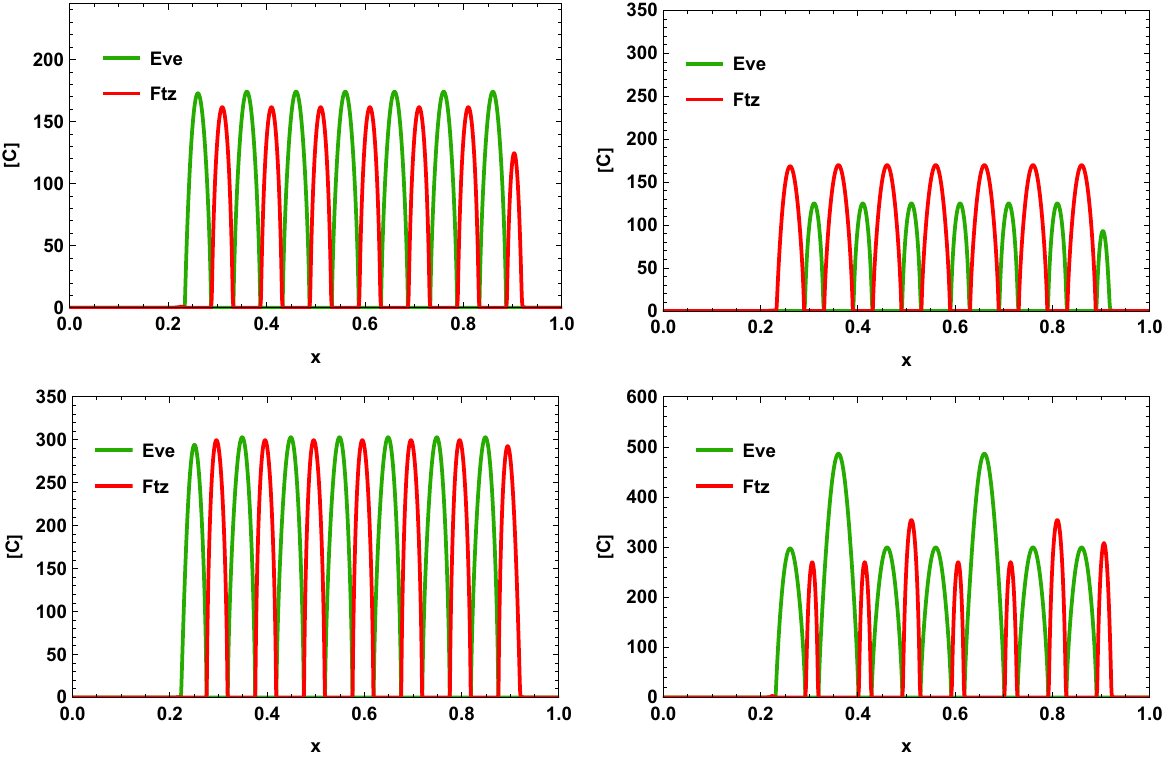}
    \caption{\small{Simulation of direct repression between \textit{Eve} and \textit{Ftz} as the main mechanism to develop the \textit{Ftz} complementary pattern. Three scenarios were tested. \textit{Eve}'s pattern is primarily formed by its five non-terminal enhancers, and \textit{Ftz}, by some unknown process, starts developing with the \textbf{(top left)} same or \textbf{(bottom left)} similar pattern to \textit{Eve}'s. Epistasis quickly overrides this mechanism and becomes the main process, creating the two complementary patterns. \textbf{(bottom right)} \textit{Ftz} starts to be produced from its stripe-specific enhancers, and then epistasis overrides and becomes the main mechanism. We assume the \textit{ftz} enhancers react to \textit{eve's} enhancers parameters. For example, we could have as an initial condition the 1+5 enhancer giving rise to a stripe overlapping \textit{Eve}'s stripe 1, and the 2+7 and 3+6 enhancers originating stripes coinciding with \textit{Eve} stripe pairs 3+7 and 4+6. Parameter values: \textbf{(top left)} k$_1$ = 3.54, k$_2$ = 3.54, k$_3$ = 0.1, k$_4$ = 0.1, k$_5$ = 2.22, k$_{-5}$ = 5.00 $\times$ 10$^{-4}$, k$_6$ = 3.66, k$_{-6}$ = 5.00 $\times$ 10$^{-4}$, k$_7$ = 2.12, k$_{-7}$ = 5.00 $\times$ 10$^{-4}$, k$_8$ = 2.12, k$_{-8}$ = 5.00 $\times$ 10$^{-4}$, $\alpha_{eve}$ = 0.365, $\alpha_{ftz}$ = 0.478; \textbf{(bottom left)} k$_1$ = 3.54, k$_2$ = 3.54, k$_3$ = 0.1, k$_4$ = 0.1, k$_5$ = 2.22, k$_{-5}$ = 5.00 $\times$ 10$^{-4}$, k$_6$ = 3.66, k$_{-6}$ = 5.00 $\times$ 10$^{-4}$, k$_7$ = 2.12, k$_{-7}$ = 5.00 $\times$ 10$^{-4}$, k$_8$ = 2.12, k$_{-8}$ = 5.00 $\times$ 10$^{-4}$, $\alpha_{eve}$ = 0.340, $\alpha_{ftz}$ = 0.478; \textbf{(bottom right)} k$_1$ = 3.54, k$_2$ = 3.54, k$_3$ = 0.1, k$_4$ = 0.1, k$_5$ = 1.40, k$_{-5}$ = 5.00 $\times$ 10$^{-4}$, k$_6$ = 3.00, k$_{-6}$ = 5.00 $\times$ 10$^{-4}$, k$_7$ = 2.12, k$_{-7}$ = 5.00 $\times$ 10$^{-4}$, k$_8$ = 2.12, k$_{-8}$ = 5.00 $\times$ 10$^{-4}$,  $\alpha_{eve}$ = 0.340, $\alpha_{ftz}$ = 0.796. Suppose we reverse each protein's contribution and the rate constants responsible for direct repression between \textit{Eve} and \textit{Ftz}. In that case, we obtain swapped protein concentration profiles in space \textbf{(top right)}. In the latter case the following parameter values were considered: k$_1$ = 3.54, k$_2$ = 3.54, k$_3$ = 0.1, k$_4$ = 0.1, k$_5$ = 3.52, k$_{-5}$ = 5.00 $\times$ 10$^{-4}$, k$_6$ = 2.34, k$_{-6}$ = 5.00 $\times$ 10$^{-4}$, k$_7$ = 2.12, k$_{-7}$ = 5.00 $\times$ 10$^{-4}$, k$_8$ = 2.12, k$_{-8}$ = 5.00 $\times$ 10$^{-4}$, $\alpha_{eve}$ = 0.452, $\alpha_{ftz}$ = 0.358.}}
    \label{fig:alternative}
\end{figure*}

Modelling \textit{Eve} and \textit{Ftz} pattern development with a purely repressive gap regulating system was successful, with the two proteins demonstrating the desired complementarity. In Figure \ref{fig:full_ftz_eve}, we combine the two distributions obtained in Figures~\ref{fig:all_eve_fits} and \ref{fig:all_ftz_fits}. We provided a new mechanism for the regulation of stripe 1, with a repressive combination of five gap proteins, making us reconsider the \textit{Hb} activation proposal by \cite{Fuji}.

The final \textit{Eve} and \textit{Ftz} patterns have $\chi^2$ values of $\sqrt{\chi^2_{Eve}}=47$ and  $\sqrt{\chi^2_{Ftz}}=46$, which are dependent on the system of units of experimental data. On the other hand,  this depends on the variability of embryo lengths and the experimental conditions used to obtain each protein profile. These variabilities, although small, can introduce imprecisions in the stripe location predictions. Ideally, we would have to use profiles belonging to the same embryo to be objective.

\subsection{\textbf{An alternative \textit{Ftz} pattern formation via \textit{eve}-\textit{ftz} epistasis}}

In the models developed stripe 4 expressions cannot be explained by gap transcriptional repression. The two patterns we obtained (Figure \ref{fig:full_ftz_eve}) are not perfectly complementary. This suggests yet another level of complexity as, for example, the introduction of cross-repression between \textit{Eve} and \textit{Ftz} to adjust the precise position of the stripes and develop \textit{Ftz}'s stripe 4.  This hypothesis advanced by \cite{Frash_correct}, together with studies by  \cite{Lim2018} on the temporal dynamics of the two proteins, led us to consider the possibility of another, more straightforward mechanism that can override the stripe-specific enhancers action where \textit{Eve} and \textit{Ftz} directly repress each other's transcription. 

To test this self-repression hypothesis, we introduced the simplified kinetic mechanism
\begin{equation} \label{SEFsystem}
        \begin{array}{ll}\displaystyle 
        \text{G}_\text{e}  \mathop{\longrightarrow}^{k_{1}} \text{G}_\text{e} + \text{Eve}, & \displaystyle
        \text{G}_\text{f}  \mathop{\longrightarrow}^{k_{2}} \text{G}_\text{f} + \text{Ftz}, \\ \displaystyle
        \text{Eve} \mathop{\longrightarrow}^{k_{3}}, & \displaystyle
        \text{Ftz} \mathop{\longrightarrow}^{k_{4}}, \\ \displaystyle
        \text{Eve} + \text{G}_\text{f} \mathop{\rightleftarrows}^{k_{5}}_{k_{-5}} \text{G}_{\text{R1}}, & \displaystyle
        \text{Ftz} + \text{G}_\text{e} \mathop{\rightleftarrows}^{k_{6}}_{k_{-6}} \text{G}_{\text{R2}}, \\ \displaystyle
        \text{TR} + \text{G}_\text{e} \mathop{\rightleftarrows}^{k_{7}}_{k_{-7}} \text{G}_{\text{R3}}, & \displaystyle
        \text{TR} + \text{G}_\text{f} \mathop{\rightleftarrows}^{k_{8}}_{k_{-8}} \text{G}_{\text{R4}}
    \end{array}
\end{equation} 
and we mimicked the stripped patterns using sums of Gaussian functions, each centred in individual stripes. The terminal repression is ensured by the hypothetical terminal regulator protein \text{TR} and was mimicked using the Error function. The evolution equations of the kinetic mechanism \eqref{SEFsystem} are shown in \ref{AppB}. 

In  Figure \ref{fig:alternative}, we show the steady states obtained with the evolution equations associated with kinetic mechanism \eqref{SEFsystem} for three different biological scenarios.  In every case, this epistasis mechanism proved sufficient to produce two perfectly complementary stripe patterns.
Although successful in theory, this epistasis concept implies that the \textit{Ftz} pattern formation depends on \textit{Eve}. At first glance, this does not agree with the mutation experiments (\cite{Harding}). However, we point out that there are few experiments of this type and little information on the regulation of \textit{ftz} for a conclusive study. Moreover, we must not forget that \textit{eve} and \textit{ftz} are part of a complex gene expression network. We can conjecture a cross-repression network between the pair-rule such that other genes similar to \textit{Eve} assume its role in providing periodic input in the case of a gene mutation.

\subsection{\textbf{Modeling segment-polarity pattern formation}}

\begin{figure}[t] 
        \centering
        \includegraphics[width=0.95\columnwidth]{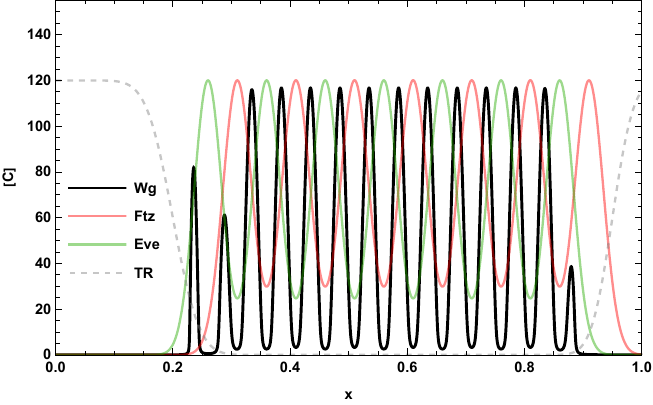}
         \includegraphics[width=0.95\columnwidth]{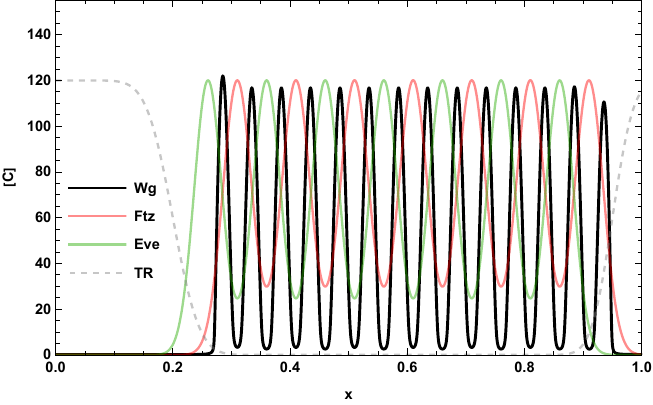}
\caption{Simulation of a segment-polarity pattern (solid black line), through the repressive action of \textit{Eve} (solid green line) and \textit{Ftz} (solid red line) in \textit{Drosophila} embryogenesis. We observe the clear formation of the expected 14 stripes. Terminal repression was introduced via an Error function, whose parameters determine the positioning of the fourteenth stripe: \textbf{(top)} before the first \textit{Eve} stripe or \textbf{(bottom)} after the last \textit{Ftz} stripe. Steady-state solutions evaluated at time $t=100$. Parameter values: \textbf{(top)}  k$_1$ = 13.65, k$_2$ = 1.00, k$_{3/4}$ = 5.00, k$_{-3/-4}$ = 10$^{-2}$, $\alpha_0$ = 1.00, $\alpha_1$ = 0.57, $\alpha_2$ = 0.60 and $\alpha_4$ = 0.15; \textbf{(bottom)}  k$_1$ = 13.65, k$_2$ = 1.00, k$_{3/4}$ = 5.00, k$_{-3/-4}$ = 10$^{-2}$, $\alpha_0$ = 1.00, $\alpha_1$ = 0.57, $\alpha_2$ = 0.60 and $\alpha_4$ = 0.42.}   
    \label{fig:wg}
\end{figure}

Our modelling strategy can be extended to infer the behaviour of other genes in the \textit{Drosophila} embryogenesis, just as the segment-polarity gene \textit{Wg}. \textit{Wg} has 14 stripes perfectly developed in the narrow stripes of cells that separate \textit{Eve} and \textit{Ftz} (\cite{Ingham}). We built a simple model 
\begin{equation}\label{SPsystem}
    \begin{array}{ll}
        \displaystyle
        \text{G}_{wg} \mathop{\longrightarrow}^{k_{1}} \text{G}_{wg} + \text{Wg}, & \displaystyle
        \text{Wg} \mathop{\longrightarrow}^{k_{2}}, \\ \displaystyle
         \text{Eve} + \text{G}_{wg} \mathop{\rightleftarrows}^{k_{3}}_{k_{-3}} \text{G}_{R1},& \displaystyle
       \text{Ftz} + \text{G}_{wg} \mathop{\rightleftarrows}^{k_{4}}_{k_{-4}} \text{G}_{R2} \\ \displaystyle
        \text{TR} + \text{G}_{wg} \mathop{\rightleftarrows}^{k_{5}}_{k_{-5}} \text{G}_{R3}. &
    \end{array}
\end{equation}
where $\text{G}_{wg}$ is the gene  for the free production of the protein \textit{Wg}, and \textit{Eve} and \textit{Ftz} are repressive regulators. 
As we don't know which proteins regulate terminal gene expression, we introduced terminal repressive action by the terminal regulator TR, whose concentration distribution throughout the embryo is simulated by an Error function.
 The patterns of \textit{Eve} and \textit{Ftz} were simulated, summing localised Gaussian functions, each centred in an individual stripe.  The corresponding evolution equations derived from the mass action law are shown in \ref{AppC}. The model results in Figure \ref{fig:wg} show the full development of the 14 stripes. By changing the parameters of the Error function, that is, its shape and the way it intersects the \textit{Eve} and \textit{Ftz} stripe patterns, we show how the fourteenth \textit{Wg} stripe can appear before the first \textit{Eve} stripe (Figure \ref{fig:wg}-(top)) or after the last \textit{Ftz} stripe (Figure \ref{fig:wg}-(bottom)). 

This demonstrates that with access to experimental data for segment-polarity expression patterns, it would be easy to analyse the possible pair-rule combinations, determine transcriptional activators and repressors for this last hierarchical level of gene regulation, and calibrate the respective patterns, regardless of the embryo in question.

\section{Conclusions}
\label{sec:concl}

We studied the gene regulatory network of \textit{Drosophila} embryogenesis, focusing on modelling and calibrating the seven-stripe patterns of the proteins \textit{Eve} and \textit{Ftz}. We modelled seven stripe-specific enhancers for \textit{eve} with the kinetic mass-action law describing transcriptional regulation. We proposed a purely repressive gap system with novel minimal regulator combinations as depicted in Figure~\ref{fig:enhancers}. Our model predictions matched the data remarkably well. They provided a complete description of \textit{Eve's} development, including the definition of the anterior and posterior \textit{Drosophila} embryo poles and the calibration of the \textit{Eve} production rate across the embryo.   Moreover,  the calibrated model reproduces gap gene mutation experiments.

We identified three gap repressive combinations driving three \textit{ftz} enhancers and successfully achieved \textit{Ftz's} pattern development calibration, except stripe 4, Figure~\ref{fig:figftz}. We explored an alternative mechanism of epistasis between \textit{eve} and \textit{ftz} that proved to be sufficient to originate two perfectly complementary patterns, having thus identified two different mechanisms for \textit{Ftz} formation.  Lastly, extended work inferring the Wingless (\textit{Wg}) fourteen stripe pattern from \textit{Eve} and \textit{Ftz} repressive transcription factors have been proposed,  clarifying the hierarchical structure of \textit{Drosphila}'s genetic expression network during early development before cellularisation.

\section*{Ackowledgements:}  We wish to express our gratitude to the two anonymous referees of the paper for their valuable comments, which significantly contributed to improving the text.

\section*{CRediT author statement} 
\textbf{Catarina Dias:} Conceptualization, Data curation, Formal analysis, Investigation, Methodology, Software, Visualization, Writing - original draft, Writing - review \& editing. 
\textbf{Rui Dil\~ao:} Conceptualization,  Formal analysis, Investigation, Methodology, Software, Supervision, Writing - review \& editing.

\section*{Declaration of interest}
The authors declare no conflict or competing interests.


\appendix 

\section{Stripe-specific \textit{eve} enhancers}\label{AppA}

\subsection{Transcriptional regulation model for stripes 1 and 2 of \textit{Eve}, and dual stripe 1+5 of \textit{Ftz}}

\begin{equation} \label{S1-2systemSup}
    \begin{array}{ll}
        \displaystyle
        \text{G} \mathop{\longrightarrow}^{k_{1}} \text{G} + \text{Eve}, &\displaystyle
        \text{Eve} \mathop{\longrightarrow}^{d},\\ \displaystyle
        \text{Kr} + \text{G} \mathop{\rightleftarrows}^{k_{3}}_{k_{-3}} \text{G}_{R1}, &\displaystyle
        \text{Kni} + \text{G}\mathop{\rightleftarrows}^{k_{4}}_{k_{-4}} \text{G}_{R2},\\ \displaystyle
        \text{Gt} + \text{G} \mathop{\rightleftarrows}^{k_{5}}_{k_{-5}} \text{G}_{R3}, &\displaystyle
        \text{Tll} + \text{G} \mathop{\rightleftarrows}^{k_{6}}_{k_{-6}} \text{G}_{R4}, \\ \displaystyle
        \text{Hkb} + \text{G} \mathop{\rightleftarrows}^{k_{7}}_{k_{-7}} \text{G}_{R5}. &
    \end{array}
\end{equation} 
In the case of \textit{Eve} we considered the gene  $\text{G}\to  \text{G}_e$ and,  in the case of \textit{Ftz}, $\text{G}\to  \text{G}_f$. The stripes 1 and 2 production models of  \textit{Eve} and the production model for dual stripe 1+5  have different rates.

The production of proteins \textit{Eve} or \textit{Ftz} described by the kinetic equations \eqref{S1-2systemSup} are
\scriptsize{ $$
\begin{array}{lcl}
    \overline{\text{Eve}}'(t) &= & -d \overline{\text{Eve}}(t) + \alpha _1 k_1 F(\cdot)\\
                  \overline{\text{Kr}}'(t)&=&k_{-3} \left[\alpha _2 \text{G}_{R1}(0)-\overline{\text{Kr}}(t)+\overline{\text{Kr}}(0)\right]-k_3 \overline{\text{Kr}}(t) F(\cdot) \\[5pt]
                          \overline{\text{Kni}}'(t)&=&k_{-4} \left[\alpha_3 \text{G}_{R2}(0)-\overline{\text{Kni}}(t)+\overline{\text{Kni}}(0)\right]-k_4 \overline{\text{Kni}}(t)F(\cdot)\\[5pt]
\overline{\text{Gt}}'(t)&=&k_{-5} \left[\alpha_4 \text{G}_{R3}(0)-\overline{\text{Gt}}(t)+\overline{\text{Gt}}(0)\right]-k_5 \overline{\text{Gt}}(t) F(\cdot)\\[5pt]
        \overline{\text{Tll}}'(t)&=&k_{-6} \left[\alpha_6 \text{G}_{R4}(0)-\overline{\text{Tll}}(t)+\overline{\text{Tll}}(0)\right]-k_6 \overline{\text{Tll}}(t) F(\cdot)\\[5pt]
        \overline{\text{Hkb}}'(t)&=&k_{-7} \left[\alpha_7 \text{G}_{R5}(0)-\overline{\text{Hkb}}(t)+\overline{\text{Hkb}}(0)\right]-k_7 \overline{\text{Hkb}}(t) F(\cdot) ,
\end{array}
$$ }
where
\scriptsize{$$
\begin{array}{ll}
F(\cdot)= & \left(\frac{\overline{\text{G}}(0)}{\alpha_0}+\frac{\overline{\text{Kr}}(t)-\overline{\text{Kr}}(0)}{\alpha_2}+\frac{\overline{\text{Kni}}(t)-\overline{\text{Kni}}(0)}{\alpha_3}+\frac{\overline{\text{Gt}}(t)-\overline{\text{Gt}}(0)}{\alpha_4}\right. \\ & \left. +\frac{\overline{\text{Tll}}(t)-\overline{\text{Tll}}(0)}{\alpha_6}+\frac{\overline{\text{Hkb}}(t)-\overline{\text{Hkb}}(0)}{\alpha_7}  \right),
\end{array}
$$}
P or  $\overline{\text{P}}=\alpha_{\text{p}}\text{P}$ represent a specific protein. The calibration parameters  for each repressor proteins are: $\alpha_0=\alpha_{G_e/G_f}$, $\alpha_1=\alpha_{eve/ftz}$, $\alpha_2=\alpha_{kr}$, $\alpha_3=\alpha_{kni}$, $\alpha_4=\alpha_{gt}$, $\alpha_5=\alpha_{hb}$, $\alpha_6=\alpha_{tll}$ and $\alpha_7=\alpha_{hkb}$. 
In all the simulations, we have assumed $\alpha_0=1$. These conditions are fixed for all the models analysed.

At each anteroposterior $x$-coordinate of the embryo, the repressive protein data are introduced at $\overline{\text{P}}(0)$. The initial conditions for each repressor complex are $\text{G}_{R1}(0)=\ldots =\text{G}_{R5}(0)=0$ and $\overline{\text{G}}(0)=250$. The initial conditions for \textit{Eve} and \textit{Ftz} models are $\overline{\text{Eve}}(0)=0$ and $\overline{\text{Ftz}}(0)=0$.

\subsection{Transcriptional regulation model for dual stripes 3+7 of \textit{Eve}, and dual stripe 2+7 and 3+6 of \textit{Ftz}}

\begin{equation} \label{S37systemSup}
    \begin{array}{ll}
        \displaystyle
        \text{G} \mathop{\longrightarrow}^{k_{1}} \text{G} + \text{Eve}, &\displaystyle
        \text{Eve} \mathop{\longrightarrow}^{d} , \\\displaystyle 
        \text{Kni} + \text{G} \mathop{\rightleftarrows}^{k_{3}}_{k_{-3}} \text{G}_{R1}, &\displaystyle
       \text{Hb} + \text{G} \mathop{\rightleftarrows}^{k_{4}}_{k_{-4}} \text{G}_{R2},\\\displaystyle 
        \text{Tll} + \text{G} \mathop{\rightleftarrows}^{k_{5}}_{k_{-5}} \text{G}_{R3}, &\displaystyle
        \text{Hkb} + \text{G} \mathop{\rightleftarrows}^{k_{6}}_{k_{-6}} \text{G}_{R4}.
    \end{array}
\end{equation}

The production of protein \textit{Eve} or \textit{Ftz} described by the kinetic equations \eqref{S37systemSup} are
$$
\begin{array}{lcl}
    \overline{\text{Eve}}'(t) &=& -d \overline{\text{Eve}}(t) + \alpha_1 k_1 F(\cdot) \\[5pt]
        \overline{\text{Kni}}'(t)&= & k_{-3} \left[\alpha_3 \text{G}_{R1}(0)-\overline{\text{Kni}}(t)+\overline{\text{Kni}}(0)\right]-k_3 \overline{\text{Kni}}(t) F(\cdot) \\[5pt]
    \overline{\text{Hb}}'(t)&= & k_{-4} \left[\alpha_5 \text{G}_{R2}(0)-\overline{\text{Hb}}(t)+\overline{\text{Hb}}(0)\right]- k_4 \overline{\text{Hb}}(t) F(\cdot) \\[5pt]
    \overline{\text{Tll}}'(t) & =&k_{-5} \left[\alpha_6 \text{G}_{R3}(0)-\overline{\text{Tll}}(t)+\overline{\text{Tll}}(0)\right]-k_5 \overline{\text{Tll}}(t) F(\cdot) \\[5pt] 
    \overline{\text{Hkb}}'(t)&= & k_{-6} \left[\alpha_7 \text{G}_{R4}(0)-\overline{\text{Hkb}}(t)+\overline{\text{Hkb}}(0)\right]-k_6 \overline{\text{Hkb}}(t) F(\cdot) ,
\end{array}
$$
where
\scriptsize{$$
\begin{array}{ll}
F(\cdot)= &  \left(\frac{\overline{\text{G}}(0)}{\alpha_0}+ \frac{\overline{\text{Kni}}(t) - \overline{\text{Kni}}(0)}{\alpha_3}+\frac{\overline{\text{Hb}}(t)-\overline{\text{Hb}}(0)}{\alpha_5} + \frac{\overline{\text{Tll}}(t)-\overline{\text{Tll}}(0)}{\alpha_6}\right. \\ & \left. + \frac{\overline{\text{Hkb}}(t)-\overline{\text{Hkb}}(0)}{\alpha_7}    \right).
\end{array}
$$}

\subsection{Transcriptional regulation model for dual stripes 4+6 of \textit{Eve}}

\begin{equation}  \label{S46systemSup}
 \begin{array}{ll}
        \displaystyle
        \text{G}_e \mathop{\longrightarrow}^{k_{1}} \text{G}_e + \text{Eve}, & \displaystyle
        \text{Eve} \mathop{\longrightarrow}^{d}, \\ \displaystyle
        \text{Kni} + \text{G}_e \mathop{\rightleftarrows}^{k_{3}}_{k_{-3}} \text{G}_{R1}, & \displaystyle
        \text{Gt} + \text{G}_e \mathop{\rightleftarrows}^{k_{4}}_{k_{-4}} \text{G}_{R2}, \\ \displaystyle
        \text{Hb} + \text{G}_e \mathop{\rightleftarrows}^{k_{5}}_{k_{-5}} \text{G}_{R3},&\displaystyle
        \text{Tll} + \text{G}_e \mathop{\rightleftarrows}^{k_{6}}_{k_{-6}} \text{G}_{R4}, \\ \displaystyle
        \text{Hkb} + \text{G}_e \mathop{\rightleftarrows}^{k_{7}}_{k_{-7}} \text{G}_{R5}. &
    \end{array}
\end{equation}

The production of protein \textit{Eve} described by the kinetic equations \eqref{S46systemSup} is
$$
\begin{array}{lcl}
    \overline{\text{Eve}}'(t) &= & -d \overline{\text{Eve}}(t) + \alpha _1 k_1 F(\cdot) \\
            \overline{\text{Kni}}'(t)&= &k_{-3} \left[\alpha_3 \text{G}_{R1}(0)-\overline{\text{Kni}}(t)+\overline{\text{Kni}}(0)\right]-k_3 \overline{\text{Kni}}(t) F(\cdot)  \\[5pt]
        \overline{\text{Gt}}'(t)&= & k_{-4} \left[\alpha_4 \text{G}_{R2}(0)-\overline{\text{Gt}}(t)+\overline{\text{Gt}}(0)\right]-k_4 \overline{\text{Gt}}(t) F(\cdot) \\[5pt]
        \overline{\text{Hb}}'(t)&= &k_{-5} \left[\alpha_5 \text{G}_{R3}(0)-\overline{\text{Hb}}(t)+\overline{\text{Hb}}(0)\right]-k_5 \overline{\text{Hb}}(t) F(\cdot) \\[5pt]
       \overline{\text{Tll}}'(t)&=&k_{-6} \left[\alpha_6 \text{G}_{R4}(0)-\overline{\text{Tll}}(t)+\overline{\text{Tll}}(0)\right]-k_6 \overline{\text{Tll}}(t) F(\cdot)\\[5pt]
        \overline{\text{Hkb}}'(t)&=&k_{-7} \left[\alpha_7 \text{G}_{R5}(0)-\overline{\text{Hkb}}(t)+\overline{\text{Hkb}}(0)\right]-k_7 \overline{\text{Hkb}}(t) F(\cdot) ,
\end{array}
$$
where
\scriptsize{$$
\begin{array}{ll}
F(\cdot)=& \left(\frac{\overline{\text{G}}_e(0)}{\alpha_0}+\frac{\overline{\text{Kni}}(t)-\overline{\text{Kni}}(0)}{\alpha_3}+\frac{\overline{\text{Gt}}(t)-\overline{\text{Gt}}(0)}{\alpha_4}+\frac{\overline{\text{Hb}}(t)-\overline{\text{Hb}}(0)}{\alpha_5}\right. \\ &\left.+\frac{\overline{\text{Tll}}(t)-\overline{\text{Tll}}(0)}{\alpha_6}+ \frac{\overline{\text{Hkb}}(t)-\overline{\text{Hkb}}(0)}{\alpha_7}
\right).
\end{array}
$$}

\subsection{Transcriptional regulation model for stripe 5 of \textit{Eve}}

\begin{equation}  \label{S5system2Sup}
    \begin{array}{ll}
        \displaystyle
        \text{G}_e \mathop{\longrightarrow}^{k_{1}} \text{G}_e + \text{Eve}, & \displaystyle
        \text{Eve} \mathop{\longrightarrow}^{d}, \\ \displaystyle
        \text{Kr} + \text{G}_e \mathop{\rightleftarrows}^{k_{3}}_{k_{-3}} \text{G}_{R1} ,& \displaystyle
        \text{Gt} + \text{G}_e \mathop{\rightleftarrows}^{k_{4}}_{k_{-4}} \text{G}_{R2}, \\ \displaystyle
        \text{Hb} + \text{G}_e \mathop{\rightleftarrows}^{k_{5}}_{k_{-5}} \text{G}_{R3}, & \displaystyle
        \text{Tll} + \text{G}_e \mathop{\rightleftarrows}^{k_{6}}_{k_{-6}} \text{G}_{R4},\\ \displaystyle
        \text{Hkb} + \text{G}_e \mathop{\rightleftarrows}^{k_{7}}_{k_{-7}} \text{G}_{R5} .&
    \end{array}
\end{equation}

The production of protein \textit{Eve} described by the kinetic equations \eqref{S5system2Sup} is
$$
\begin{array}{lcl}
    \overline{\text{Eve}}'(t) &= & -d \overline{\text{Eve}}(t) + \alpha_1 k_1F(\cdot)\\
        \overline{\text{Kr}}'(t)&=&k_{-3}  \left[\alpha _2 \text{G}_{R1}(0)- \overline{\text{Kr}}(t)+\overline{\text{Kr}}(0)\right]- k_3 \overline{\text{Kr}}(t) F(\cdot)\\[5pt]
    \overline{\text{Gt}}'(t)&=& k_{-4} \left[\alpha_4 \text{G}_{R2}(0) -\overline{\text{Gt}}(t)+\overline{\text{Gt}}(0)\right]- k_4  \overline{\text{Gt}}(t) F(\cdot)\\  [5pt]
        \overline{\text{Hb}}'(t)&=&k_{-5} \left[\alpha_5 \text{G}_{R3}(0)- \overline{\text{Hb}}(t)+\overline{\text{Hb}}(0)\right]-k_5 \overline{\text{Hb}}(t) F(\cdot)\\[5pt]
    \overline{\text{Tll}}'(t)&=& k_{-6} \left[\alpha_6 \text{G}_{R4}(0)-\overline{\text{Tll}}(t)+\overline{\text{Tll}}(0)\right]- k_6 \overline{\text{Tll}}(t) F(\cdot)\\[5pt]
        \overline{\text{Hkb}}'(t)&=&k_{-7}  \left[\alpha_7 \text{G}_{R5}(0)- \overline{\text{Hkb}}(t)+\overline{\text{Hkb}}(0)\right]- k_7 \overline{\text{Hkb}}(t) F(\cdot),
\end{array}
$$
where
\scriptsize{$$
\begin{array}{ll}
F(\cdot)= & \left(\frac{\overline{\text{G}}_e(0)}{\alpha_0} + \frac{\overline{\text{Kr}}(t)-\overline{\text{Kr}}(0)}{\alpha_2} +\frac{\overline{\text{Gt}}(t)-\overline{\text{Gt}}(0)}{\alpha_4}+\frac{\overline{\text{Hb}}(t)-\overline{\text{Hb}}(0)}{\alpha_5}\right. \\ &\left. +\frac{\overline{\text{Tll}}(t)-\overline{\text{Tll}}(0)}{\alpha_6} +\frac{\overline{\text{Hkb}}(t)-\overline{\text{Hkb}}(0)}{\alpha_7}   \right).
\end{array}
$$}

\section{Alternative pattern formation: \textit{eve-ftz} epistasis}\label{AppB}

\begin{equation} \label{SEFsystemSup}
    \begin{array}{ll}\displaystyle 
    \text{G}_\text{e}  \mathop{\longrightarrow}^{k_{1}} \text{G}_\text{e} + \text{Eve}, & \displaystyle
    \text{G}_\text{f}  \mathop{\longrightarrow}^{k_{2}} \text{G}_\text{f} + \text{Ftz}, \\ \displaystyle
    \text{Eve} \mathop{\longrightarrow}^{k_{3}}, & \displaystyle
    \text{Ftz} \mathop{\longrightarrow}^{k_{4}}, \\ \displaystyle
    \text{Eve} + \text{G}_\text{f} \mathop{\rightleftarrows}^{k_{5}}_{k_{-5}} \text{G}_{\text{R1}}, & \displaystyle
    \text{Ftz} + \text{G}_\text{e} \mathop{\rightleftarrows}^{k_{6}}_{k_{-6}} \text{G}_{\text{R2}}, \\ \displaystyle
    \text{TR} + \text{G}_\text{e} \mathop{\rightleftarrows}^{k_{7}}_{k_{-7}} \text{G}_{\text{R3}}, & \displaystyle
    \text{TR} + \text{G}_\text{f} \mathop{\rightleftarrows}^{k_{8}}_{k_{-8}} \text{G}_{\text{R4}}.
\end{array}
\end{equation}

The production of proteins \textit{Eve} and  \textit{Ftz}  described by the kinetic equations \eqref{SEFsystemSup} is
$$
\begin{array}{lcl}
 \overline{\text{Eve}}'(t)&=& -\overline{\text{Eve}}(t) \left(k_3 + \frac{\overline{\text{G}}_f (t) k_5}{\alpha_1}\right) + \frac{\overline{\text{G}}_e (t) k_1 \alpha_2}{\alpha_0} + k_{-5}\alpha_2 F_1(\cdot) \\[5pt]
 \overline{\text{Ftz}}'(t)&=& - \overline{\text{Ftz}}(t) \left(k_4 + \frac{\overline{\text{G}}_e (t) k_6}{\alpha_0} \right) + \alpha_3 \left( k_{-6} \text{G}_{R2}(t) + \frac{\overline{G}_f(t) k_2}{\alpha_1} \right)\\[5pt]
  \overline{\text{G}}_e'(t)&=& -\overline{\text{G}}_e(t) \left( \frac{\overline{\text{Ftz}}(t) k_6}{\alpha_3} + \frac{\overline{\text{TR}}(t)k_7}{\alpha_4} \right) + k_{-6}\alpha_0 \text{G}_{R2}(t) - k_{-7}\alpha_0 F_2(\cdot)\\[5pt]
 \overline{\text{G}}_f'(t)&=& -\overline{\text{G}}_f(t) \left(\frac{\overline{\text{Eve}}(t) k_5}{\alpha_2} +\frac{\overline{\text{TR}}(t) k_8}{\alpha_4}\right) + k_{-5} \alpha_1 F_1(\cdot) + k_{-8} \alpha_1 F_3(\cdot)\\[5pt]
 \overline{\text{TR}}'(t)&=& - \overline{\text{TR}}(t) \left(\frac{\overline{\text{G}}_e(t) k_{7}}{\alpha_0} + \frac{\overline{\text{G}}_f(t) k_8}{\alpha_1}\right) -k_{-7} \alpha_4 F_2(\cdot)+ k_{-8} \alpha_4 F_3(\cdot)\\[5pt]
 \text{G}_{R2}'(t)&=& -\text{G}_{R2}(t) k_{-6} + \frac{\overline{\text{Ftz}}(t)\overline{G}_e(t) k_6}{\alpha_0 \alpha_3},
\end{array}
$$
where
\scriptsize{$$
\begin{array}{lcl}
F_1(\cdot)&=& \left(\text{G}_{R1}(0) + \text{G}_{R2}(0)-\text{G}_{R2}(t) + \frac{\overline{\text{G}}_e(0) - \overline{\text{G}}_e(t)}{\alpha_0} + \frac{\overline{\text{G}}_f(0) - \overline{\text{G}}_f(t)}{\alpha_1}\right. \\&&+ \left. \frac{\overline{\text{TR}}(t) - \overline{\text{TR}}(0)}{\alpha_4} \right)\\[5pt]
F_2(\cdot)&=& \left(\text{G}_{R2}(0) - \text{G}_{R3}(0)-\text{G}_{R2}(t) + \frac{\overline{\text{G}}_f(0) - \overline{\text{G}}_f(t)}{\alpha_1} + \frac{\overline{\text{TR}}(t) - \overline{\text{TR}}(0)}{\alpha_4} \right)\\ [5pt]
F_3(\cdot)&=& \left(\text{G}_{R2}(0) + \text{G}_{R4}(0)-\text{G}_{R2}(t) + \frac{\overline{\text{G}}_f(0) - \overline{\text{G}}_f(t)}{\alpha_1} \right).
\end{array} 
$$}

The calibration parameter parameters are: $\alpha_0=\alpha_{G_e}=1$, $\alpha_1=\alpha_{G_f}=1$, $\alpha_2=\alpha_{eve}$, $\alpha_3=\alpha_{ftz}$, $\alpha_4=\alpha_{tr}$. 

\section{Segment-Polarity pattern formation}\label{AppC}

\begin{equation}\label{SPsystemSup}
    \begin{array}{ll}
        \displaystyle
        \text{G}_{wg} \mathop{\longrightarrow}^{k_{1}} \text{G}_{wg} + \text{Wg}, & \displaystyle
        \text{Wg} \mathop{\longrightarrow}^{k_{2}}, \\ \displaystyle
        \text{Eve} +  \text{G}_{wg} \mathop{\rightleftarrows}^{k_{3}}_{k_{-3}}  \text{G}_{R1},& \displaystyle
         \text{Ftz} +  \text{G}_{wg} \mathop{\rightleftarrows}^{k_{4}}_{k_{-4}}  \text{G}_{R2} \\ \displaystyle
        \text{TR} +  \text{G}_{wg} \mathop{\rightleftarrows}^{k_{5}}_{k_{-5}}  \text{G}_{R3}. &
    \end{array}
\end{equation}

The production of proteins \textit{Eve},  \textit{Ftz}  and \textit{Wg} described by the kinetic equations \eqref{SPsystemSup} is
$$
\begin{array}{lcl}
    \overline{\text{Eve}}'(t) &=& k_{-3}   \left[\alpha_1 \text{G}_{R1}(0)-\overline{\text{Eve}}(t)+\overline{\text{Eve}}(0)\right]  
 + k_3 \overline{\text{Eve}}(t) F(\cdot)\\[5pt] \displaystyle
    \overline{\text{Ftz}}'(t)&=& k_{-4}   \left[\alpha_2 \text{G}_{R2}(0)-\overline{\text{Ftz}}(t)+\overline{\text{Ftz}}(0)\right]  
    +k_4 \overline{\text{Ftz}}(t) F(\cdot)\\[5pt] \displaystyle
    \overline{\text{TR}}'(t)&=&  k_{-5}   \left[\alpha_3 \text{G}_{R3}(0)-\overline{\text{TR}}(t)+\overline{\text{TR}}(0)\right] - k_{5} \overline{\text{TR}}(t) F(\cdot)\\[5pt] \displaystyle
    \overline{\text{Wg}}'(t)&=&\alpha_4 k_1 F(\cdot)-k_2 \overline{\text{Wg}}(t), 
\end{array}
$$
where
$$
F(\cdot)= \left(\frac{\overline{\text{Eve}}(0)-\overline{\text{Eve}}(t)}{\alpha_1}+\frac{\overline{\text{Ftz}}(0)-\overline{\text{Ftz}}(t)}{\alpha_2} +\frac{\overline{\text{TR}}(0)-\overline{\text{TR}}(t)}{\alpha_3} -\frac{\overline{\text{G}}_{wg}(0)}{\alpha_0}\right).
$$

The calibration parameter parameter are: $\alpha_0=\alpha_{G_{wg}}=1$, $\alpha_1=\alpha_{eve}$, $\alpha_2=\alpha_{ftz}$, $\alpha_3=\alpha_{tr}$, $\alpha_4=\alpha_{wg}=1$. 


\begin{thebibliography}{10}

\bibitem[Akam, 1997]{akam}  Akam, M., (1987), The molecular basis for metameric pattern in the Drosophila embryo, Development 10, 1-22.

\bibitem[Alberts \textit{et al.}, 2022]{Alber} Alberts, B., Johnson, A., Lewis, J., Morgan, D., Raff, M., Roberts, K., Walter, P., (2022), Molecular Biology of the Cell, 7th edition, Garland Science, New York.

\bibitem[Alves  \textit{et al.}, 2005]{ALVES2005429}  Alves, F., Dil\~ao, R., (2005),  A simple framework to describe the regulation of gene expression in prokaryotes, Comptes Rendus Biologies, 328, 429-444, doi: 10.1016/j.crvi.2005.01.009.

\bibitem[Alves  \textit{et al.}, 2006]{aldi2006}  Alves, F., Dil\~ao, R., (2006),  Modeling segmental patterning in Drosophila: Maternal and gap genes, J. Theo. Biol. 241, 342-359, doi: 10.1016/j.jtbi.2005.11.034.

\bibitem[Alves-Pires \textit{et al.}, 1998]{AP} Alves-Pires, R., Dil\~ao, R., Neves, H., Parreira, L.,  Sainhas, J., (1998), Anisotropy- free Laplacian filters, contour detection, and 3D image reconstruction for confocal microscopy imaging, In F. Muge, M. Piedade and J. C. Pinto (ed.), Proceedings of the RecPad98, 10th Portuguese Conference on Pattern Recognition, CVRM/IST, pp. 247-251,  ISBN: 972-97711- 0-3.

\bibitem[Casanova, 1990]{Cas}  Casanova, J., (1990), Pattern formation under the control of the terminal system in the Drosophila embryo, Development 110,  621-628.

\bibitem[Cicin-Sain \textit{et al.}, 2014]{SuperFly}   Cicin-Sain, D., Pulido, A. H., Crombach,  A.,  Wotton, K. R., Jim\'enez-Guri, E., Taly, J.-F., Roma, G.,  Jaeger, J., (2014), SuperFly: a comparative database for quantified spatio-temporal gene expression patterns in early dipteran embryos, Nucleic Acids Research 43, D751-D755, doi: 10.1093/nar/gku1142.

\bibitem[Clyde  \textit{et al.}, 2003]{Clyde}  Clyde, D.,   Corado, M.,  Wu, X., (2003,)  A self-organizing system of repressor gradients establishes segmental complexity in Drosophila, Nature 426, 849-853, doi: 10.1038/nature02189.

\bibitem[Crocker  \textit{et al.}, 2016]{Crocker2016QuantitativelyPC}  Crocker, J.,  Ilsley, G. R.,   Stern, D. L., (2016), Quantitatively predictable control of Drosophila transcriptional enhancers in vivo with engineered transcription factors, Nature Genetics 48, 292-298.

\bibitem[Dil\~ao \textit{et al.}, 2010]{kinetics}  Dil\~ao, R., Muraro, D., (2010),  A Software Tool to Model Genetic Regulatory Networks. Applications to the Modeling of Threshold Phenomena and of Spatial Patterning in Drosophila, PLOS ONE 5, 1-10, doi: 10.1371/journal.pone.0010743.

\bibitem[Dil\~ao \textit{et al.}, 2010b]{DiMu} Dil\~ao, R., Muraro, D., (2010b), Calibration and validation of a genetic regulatory network model describing the production of the protein Hunchback in Drosophila early development, Comptes Rendus Biologies 333, 779-788, doi:10.1016/j.crvi.2010.09.003.

\bibitem[Dil\~ao \textit{et al.}, 2010c]{DiMu2} Dil\~ao, R., Muraro, D., (2010c), mRNA diffusion explains protein gradients in Drosophila early development, Journal of Theoretical Biology  264, 847-853, doi:10.1016/j.jtbi.2010.03.012.

\bibitem[Dil\~ao, 2014]{Di14} Dil\~ao, R., (2014), Bicoid mRNA diffusion as a mechanism of morphogenesis in Drosophila early development, Comptes Rendus Biologies 337, 679-682, doi: 10.1016/j.crvi.2014.09.004.

\bibitem[Frash  \textit{et al.}, 1987]{Frash_correct}  Frash, M., Levine, M., (1987),  Complementary patterns of \textit{even-skipped} and \textit{fushi-tarazu} expression involve their differential regulation by a common set of segmentation genes in \textit{Drosophila}, Genes \& Development 1(9),  981-995,  doi: 10.1101/gad.1.9.981.

\bibitem[Fujioka  \textit{et al.}, 1999]{Fuji}   Fujioka, M.,  Emi-Sarker, Y., Yusibova,  G. L.,  Goto, T.,  Jaynes, J. B., (1999),  Analysis of an even-skipped rescue transgene reveals both composite and discrete neuronal and early blastoderm enhancers, and multi-stripe positioning by gap gene repressor gradients, Development 126, 2527-2538.

\bibitem[Harding \textit{et al.}, 1986]{Harding}  Harding,K.,   Rushlow, C.,  Doyle, H. J.,  Hoey, T.,  Levine, M., (1986), Cross-Regulatory Interactions Among Pair-Rule Genes in \textit{Drosophila}, Science 23, 953-959, doi: 10.1126/science.3755551.

\bibitem[Ilsley  \textit{et al.}, 2013]{Ilsley2013} Ilsley, G. R.,  Fisher, J.,  Apweiler, R.,  DePace, A. H.,    Luscombe, N. M., (2013), Cellular resolution models for even skipped regulation in the entire Drosophila embryo, eLife 2, e00522,  doi: 10.7554/eLife.00522.

\bibitem[Ingham \textit{et al.}, 1988]{Ingham}  Ingham, P.,   Baker, N.,  Martinez-Arias, A., (1988), Regulation of segment polarity genes in the \textit{Drosophila} blastoderm by \textit{fushi tarazu} and \textit{even skipped}, Nature 331, 73-75, doi: 10.1038/331073a0.

\bibitem[Janssens  \textit{et al.}, 2013]{Janssens2013} H. Janssens, Lack of tailless leads to an increase in expression variability in Drosophila embryos, Developmental Biology, vol. 377 (2013) 305-317.

\bibitem[Kozlov  \textit{et al.}, 2000]{Kozlov}   Kozlov, K.,  Myasnikova, E., Reinitz, J., Kosman, D., (2000),   Method for spatial registration of the expression patterns of Drosophila segmentation genes using wavelets, Computational Technologies 5, 112-119.

\bibitem[Lim \textit{et al.}, 2018]{Lim2018}   Lim, B.,   Fukaya, T.,  Heist, T.,   Levine, M., (2018),  Temporal dynamics of pair-rule stripes in living Drosophila embryos, PNAS 115, 8376-8381, doi: 10.1073/pnas.1810430115.

\bibitem[Myasnikova  \textit{et al.}, 2001]{Myasnikova} Myasnikova, E.,   Samsonova, A.,  Kozlov, K.,  Samsonova, M.,   Reinitz, J., (2001),  Registration of the expression patterns of Drosophila segmentation genes by two independent methods, Bioinformatics 17,  3-12,   doi: 10.1038/nrg1724.

\bibitem[N\"usslein-Volhard, 2006]{Nuss} N\"usslein-Volhard, C.,  (2006), Coming to life: how genes drive development, Yale University Press.

\bibitem[Pisarev  \textit{et al.}, 2008]{Pisarev}   Pisarev, A.,  Poustelnikova, E.,  Samsonova, M. G.,    Reinitz, J., (2008),  FlyEx, the quantitative atlas on segmentation gene expression at cellular resolution, Nucleic Acids Research 37, D560 - D566.

\bibitem[Poustelnikova  \textit{et al.}, 2004]{Poustelnikova}  Poustelnikova,E.,   Pisarev, A.,   Blagov, M.,   Samsonova, M.,    Reinitz, J., (2004),  A database for management of gene expression data in situ, Bioinformatics 20, 2212-2221.

\bibitem[Rivera-Pomar  \textit{et al.}, 1996]{Riv}  Rivera-Pomar, R.,  J\"ackle, H., (1996), From gradients to stripes in Drosophila embryogenesis: filling in the gaps. Trends Genet. 12, 478-483.

\bibitem[Schroeder \textit{et al.}, 2011]{Schroeder2}  Schroeder, M. D.,   Greer, C.,  Gaul, U., (2011),  How to make stripes: deciphering the transition from non-periodic to periodic patterns in Drosophila segmentation, Development 138, 3067-3078, doi: 10.1242/dev.062141.

\bibitem[Simpson-Brose  \textit{et al.}, 1994]{SIMPSONBROSE1994855}   Simpson-Brose, M.,  Treisman, J.,   Desplan, C., (1994),  Synergy between the hunchback and bicoid morphogens is required for anterior patterning in Drosophila, Cell 78, 855-865, doi: 10.1016/S0092-8674(94)90622-X.

\bibitem[Small  \textit{et al.}, 1991]{Small1991}   Small, S.,   Kraut, R.,  Hoey, T.,  Warrior, R.,  Levine, M., (1991),  Transcriptional regulation of a pair-rule stripe in Drosophila, Genes and Development 5, 827-839, doi: 10.1101/gad.5.5.827.

\bibitem[Small  \textit{et al.}, 1992]{Small1992}   Small, S.,  Blair, A.,   Levine, M., (1992),  Regulation of even-skipped stripe 2 in the Drosophila embryo, The EMBO Journal 1, 4047-4057.

\bibitem[Small  \textit{et al.}, 1996]{SMALL1996314}  Small, S.,  Blair, A.,   Levine, M., (1996),  Regulation of Two Pair-Rule Stripes by a Single Enhancer in the Drosophila Embryo, Developmental Biology 175, 314-324, doi: 10.1006/dbio.1996.0117.

\bibitem[Sonneveld \textit{et al.}, 2020]{Sonneveld2020}   Sonneveld, S., Verhagen, B. M. P.,  Tanenbaum, M. E., (2020),  Heterogeneity in mRNA translation, Trends in Cell Biology 30, 606-618, doi: 10.1016/j.tcb.2020.04.008.

\bibitem[Stanojevic  \textit{et al.}, 1991]{Stano}  Stanojevic, D.,  Small, S.,   Levine, M., (1991),  Regulation of a Segmentation Stripe by Overlapping Activators and Repressors in the \textit{Drosophila} Embryo, Science 254, 1385-1387, doi: 10.1126/science.1683715.

\bibitem[Surkova \textit{et al.}, 2008]{SURKOVA2008844}   Surkova, S.,  Kosman, D.,  Kozlov, K.,  Manu,  Myasnikova, E.,  Samsonova, A. A.,  Spirov, A.,   Vanario-Alonso, C. E.,   Samsonova, M.,   Reinitz, J., (2008),  Characterization of the Drosophila segment determination morphome,   Developmental Biology 313, 844-862, doi: 10.1016/j.ydbio.2007.10.037.

\bibitem[Surkova  \textit{et al.}, 2013]{Surkova2013}   Surkova, S.,   Golubkova, E.,  Manu,  Panok, L.,   Mamon, L.,   Reinitz, J.,    Samsonova, M., (2013),  Quantitative dynamics and increased variability of segmentation gene expression in the Drosophila krüppel and knirps mutants, Developmental Biology 376, 99-112, doi: 10.1016/j.ydbio.2013.01.008.

\bibitem[Vincent \textit{et al.}, 2018]{Vincent2018HunchbackIC}  Vincent, B. J.,   Staller, M. V.,  L{\'o}pez-Rivera, F.,   Bragdon, M. D. J.,  Pym, E. C. G.,  Biette, K. M.,  Wunderlich, Z.,  Harden, T. T., Estrada, J.,  DePace,  A. H., (2018), Hunchback is counter-repressed to regulate even-skipped stripe 2 expression in Drosophila embryos, PLoS Genetics 14(9), e1007644, doi: 10.1371/journal.pgen.1007644.

\bibitem[Weigel \textit{et al.}, 1990]{Wei}  Weigel, D.,   Jurgens, G.,   Klingler, M.,  Jackle, H., (1990), Two gap genes mediate maternal terminal pattern information in Drosophila, Science 248, 495-498.

\bibitem[Winata \textit{et al.}, 2018]{Win} Winata, C. L., Korzh, V., (2018), The translational regulation of maternal mRNAs in time and space, FEBS Lett.  592(17): 3007-3023.

\end{thebibliography}
\end{document}